\documentclass[12pt,a4paper,final]{iopart}

\usepackage{lineno}	
\usepackage{setspace}	
\usepackage{graphicx}
\usepackage{color}
\usepackage{xparse} 
\usepackage{mdframed} 
\usepackage{lipsum} 
\usepackage{xcolor}
\usepackage{makeidx}
\usepackage{fancyhdr}	

\usepackage[breaklinks=true,colorlinks=true,linkcolor=blue,urlcolor=blue,citecolor=blue]{hyperref}
\usepackage[noabbrev, nameinlink, capitalize]{cleveref}      

\usepackage[ngerman,english]{babel}
\usepackage{xspace}
\usepackage{parskip}
\usepackage[singlelinecheck=false, aboveskip=-3pt]{subcaption}
\usepackage{textcomp}
\usepackage{gensymb}
\usepackage{siunitx}
\usepackage{iopams}
\usepackage{booktabs}
\usepackage[inline]{enumitem} 

\usepackage[margin=2.2cm,includehead,includefoot]{geometry}

\usepackage{todonotes}

\DeclareGraphicsExtensions{.eps}

\usepackage{filecontents}

\definecolor{graylight}{cmyk}{.30,0,0,.67} 
\newmdenv[ 
  linecolor=graylight,
  topline=false,
  bottomline=false,
  rightline=false,
  skipabove=\topsep,
  skipbelow=\topsep
]{leftrule}

\NewDocumentEnvironment{remark}{O{\textbf{Remark:}}} 
{\begin{leftrule}\noindent\textcolor{graylight}{#1}\par}
{\end{leftrule}}

\newcommand{\testbox}[1]{#1} 

\newcommand*{\bkg}{background\xspace}

\newcommand{\HBT}{Hanbury Brown and Twiss\xspace}
\newcommand{\hbt}{HBT\xspace}

\DeclareSIUnit\lines{grooves}
\DeclareSIUnit\grooves{grooves}
\DeclareSIUnit\cps{cps}
\DeclareSIUnit\carat{carat}
\DeclareSIUnit\ct{carat}

\newcommand{\zpl}{zero-phonon-line\xspace}

\newcommand{\ZPL}{ZPL\xspace}
\newcommand{\ZPLs}{ZPLs\xspace}
\newcommand{\Zpl}{Zero-phonon-line\xspace}

\newcommand{\pl}{photoluminescence\xspace}
\newcommand{\cwl}{center wavelength\xspace}
\newcommand{\cwls}{center wavelengths\xspace}
\newcommand{\wl}{wavelength\xspace}
\newcommand{\wls}{wavelengths\xspace}
\newcommand{\lw}{linewidth\xspace}
\newcommand{\lws}{linewidths\xspace}
\newcommand{\gt}{\ensuremath{g^{(2)}}\xspace}

\newcommand{\gtz}{\ensuremath{g^{(2)}(0)}\xspace}
\newcommand{\db}{Debye-Waller\xspace}
\newcommand{\hr}{Huang-Rhys\xspace}

\newcommand{\nd}{nanodiamond\xspace}
\newcommand{\nds}{nanodiamonds\xspace}
\newcommand{\Nd}{Nanodiamond\xspace}

\newcommand{\basd}{bead-assisted sonic disintegration\xspace}
\newcommand{\cvd}{chemical vapor deposition\xspace}

\newcommand{\CVD}{CVD\xspace}
\newcommand{\hpht}{high-pressure, high-temperature\xspace}

\newcommand{\sivc}{silicon-vacancy center\xspace}
\newcommand{\siv}{SiV center\xspace}

\newcommand{\sivs}{SiV centers\xspace}

\renewcommand{\si}{silicon\xspace}

\newcommand{\ir}{iridium\xspace}

\newcommand{\ox}{oxidation\xspace}

\newcommand{\vl}{group V\xspace}
\newcommand{\hl}{group H\xspace}
\newcommand{\Hl}{Group H\xspace}

\newcommand{\embroad}{emitter V1\xspace}
\newcommand{\emnarrow}{emitter H1\xspace}

\newcommand{\insituF}{insitu50\xspace}
\newcommand{\insituS}{insitu70\xspace}
\newcommand{\insituSn}{insitu70n\xspace}
\newcommand{\insituSo}{insitu70o\xspace}
\newcommand{\insituH}{insitu100\xspace}
\newcommand{\insituHao}{insitu100ao\xspace}
\newcommand{\implantedTao}{implanted250ao\xspace}

\begin{document}

	\title[]{Strongly inhomogeneous distribution of spectral properties of silicon-vacancy color centers in nanodiamonds}

	\author{Sarah Lindner$^{1}$, Alexander Bommer$^{1}$, Andreas Muzha$^{2}$, Anke Krueger$^{2}$, Laia Gines$^{3}$, Soumen Mandal$^{3}$, Oliver Williams$^{3}$, Elisa Londero$^{4}$, Adam Gali$^{4,5}$, Christoph Becher$^{1}$}
	\address{$^1$Universit\"at des Saarlandes, Fachrichtung Physik, Campus E2.6, 66123 Saarbr\"ucken, Germany}
	\address{$^2$Institut f\"ur Organische Chemie, Universit\"at W\"urzburg, Am Hubland, D-97074 W\"urzburg }
	\address{$^3$School of Engineering, Cardiff University, Newport Road, Cardiff CF24 3AA, Wales, United Kingdom}
	\address{$^4$Wigner Research Centre for Physics, Institute for Solid State Physics and Optics, Hungarian Academy of Sciences, PO Box 49, H-1525, Hungary}
    \address{$^5$Department of Atomic Physics, Budapest University of Technology and Economics, Budafoki \'ut 8., H-1111, Hungary}
	\ead{christoph.becher@physik.uni-saarland.de}

	\begin{abstract}
		
		The silicon-vacancy (SiV) color center in diamond is a solid-state single photon emitter and spin quantum bit suited as a component in quantum devices.	
		Here, we show that the \siv in nanodiamond exhibits a strongly inhomogeneous distribution with regard to the \cwls and \lws of the \zpl (\ZPL) emission at room temperature.
We find that the SiV centers separate in two clusters: one group exhibits \ZPLs with \cwls within a narrow range $\approx$ \SIrange{730}{742}{nm} and broad \lws between \SIlist{5; 17}{nm}, whereas the second group
comprises a very broad distribution of \cwls between \SIlist{715;835}{nm}, but narrow \lws from below \SI{1}{nm} up to \SI{4}{nm}.
	  	Supported by \emph{ab initio} Kohn-Sham density functional theory calculations we show that the ZPL shifts of the first group are consistently explained by strain in the diamond lattice.
	  	Further, we suggest, that the second group showing the strongly inhomogeneous distribution of \cwls might be comprised of modified \sivs.
	  	Whereas single photon emission is demonstrated for SiV centers of both clusters, we show that emitters from different clusters show different spectroscopic features such as variations of the phonon sideband spectra and different blinking dynamics.

	\end{abstract}	\label{abstract}

	\section{Introduction}	\label{sec::introduction}

		Recently, the negatively charged \sivc (\siv) has received growing interest in the fields of quantum communication and quantum information due to its favorable spectral properties and optically accessible spin states \cite{Becker2018, Becker2017,Sukachev2017,Rogers2014a}.
		In particular, SiV centers in low strain bulk diamond have been shown to exhibit close to Fourier-limited \lws \cite{Rogers2014,Arend2016a,Schroder2017a} and emission of indistinguishable photons \cite{Sipahigil2014} from two distinct emitters, a prerequisite for many applications in quantum technologies \cite{Sipahigil2016}.
		As the spin coherence time of SiV centers in bulk diamond is limited by phonon-induced decoherence processes down to liquid-Helium temperatures \cite{Becker2018,Sukachev2017,Jahnke2015a}, diamond nanostructures or nanodiamonds might provide an advantage based on phonon confinement effects modifying the phonon density of states.
		Additionally, some applications require diamonds of the size of a few ten nanometers.
		Examples include applications as fluorescence markers \cite{Baker2010,Merson2013} or the implementation of \sivs in photonic structures such as microcavities \cite{Albrecht2014a,Benedikter2017a} or optical antennas \cite{Alaverdyan2011,Schietinger2009}.
		\\
		In light of such applications, \nds hosting color centers provide a significant advantage:
		Suitable \sivs can be spectroscopically preselected according to desired properties and can then be relocated to be used in the target structure using pick-and-place techniques \cite{Schell2011}.
		Previous research demonstrated the production of nanodiamonds including \sivs produced by \cvd (CVD) \cite{Neu2011b}, \hpht (HPHT) \cite{Davydov2014} and \basd (BASD) \cite{Neu2011a} processes.
		Although several studies \cite{Jantzen2016, Tran2017} have indicated nearly Fourier-limited \lws of \sivs in nanodiamonds, in general the spectral properties of SiV centers are strongly dependent on strain in the diamond lattice \cite{Neu2011,Neu2013,Grudinkin2016}.
		\\
	  	In this study we investigate the properties of \sivs in \nds of sizes of \SIrange{50}{100}{nm} produced in a wet-milling process in a vibrational mill.
	  	Starting material was a CVD grown diamond film with \textit{in-situ} incorporated SiV defects.
		The advantage of wet-milled \nds with \textit{in-situ} incorporated \sivs lies in its high production rate of nanodiamonds, yielding a perfect candidate for preselection of spectral properties and consecutive implementation in target applications.
		\\
		The fluorescence spectra of the \sivs show that both the center wavelength of the \ZPL as well as the \lw of the \zpl vary strongly among different diamonds.
		Our measurements show a strong correlation between the center wavelength of the \zpl and the corresponding \lws, resulting in a previously unreported bimodal distribution.
		We find single photon emission from these \sivs over the whole range of \zpl positions and \lws, although with a large variation in photon emission rate and fluorescence stability.

		\section{Methods}	\label{sec::methods}

			\begin{figure}[tp]
				\begin{subfigure}[t]{ 0.31\linewidth}
					\centering
					\testbox{\includegraphics[trim = 0 0 0 0,  clip= true, height=5cm]{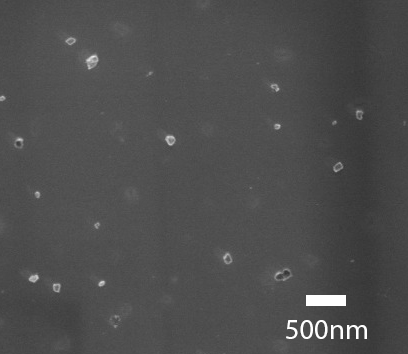}}
					\caption{}\label{subfig::sem}
				\end{subfigure}
				\hfill
				\begin{subfigure}[t]{ 0.31\linewidth}
					\centering
					\testbox{\includegraphics[trim = 0 0 0 0,  clip= true, height=5cm]{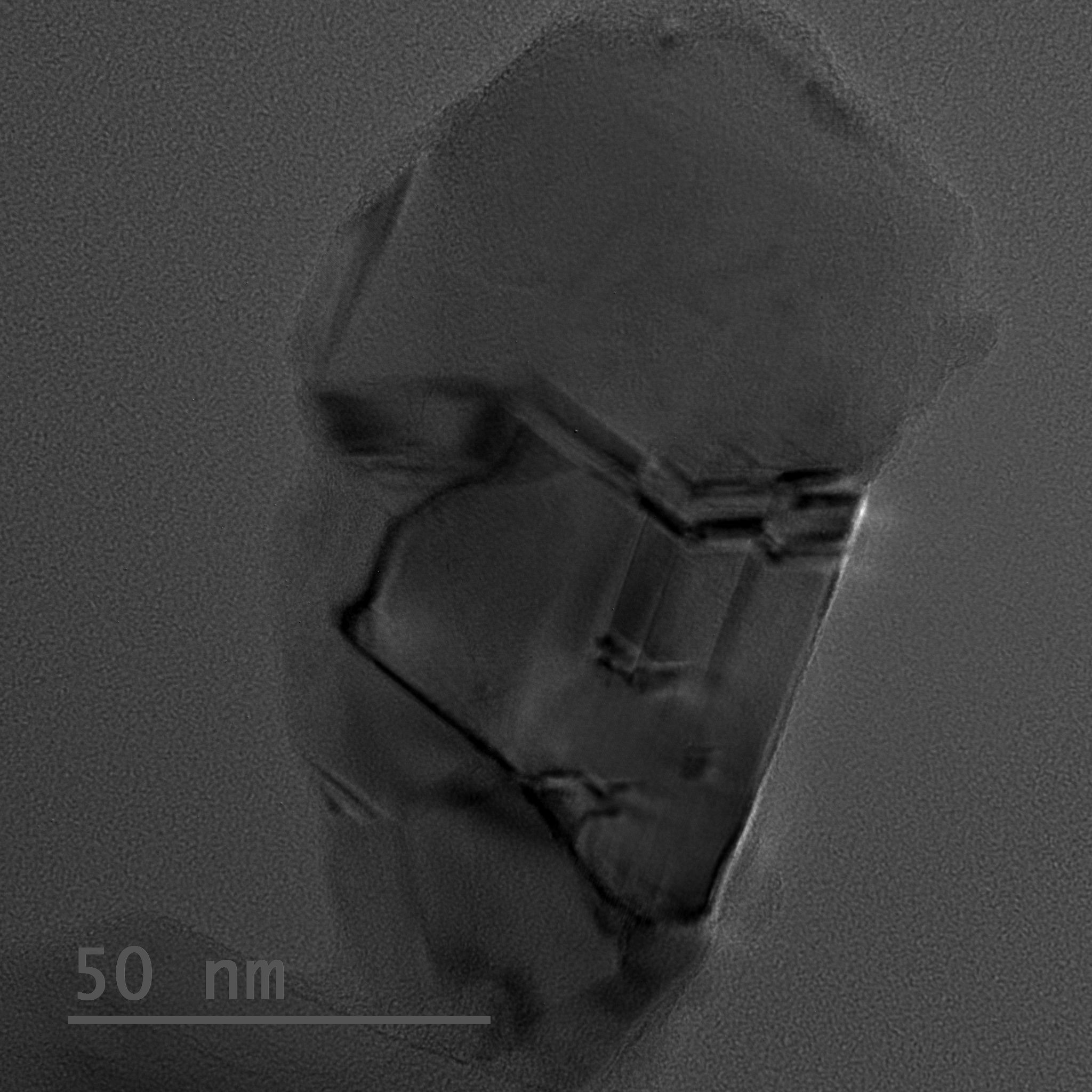}}
					\caption{}\label{subfig::tem}
					\end{subfigure}
				\begin{subfigure}[t]{ 0.31\linewidth}
					\centering
					\testbox{\includegraphics[trim = 0 0 0 0,  clip = true, height=5cm]{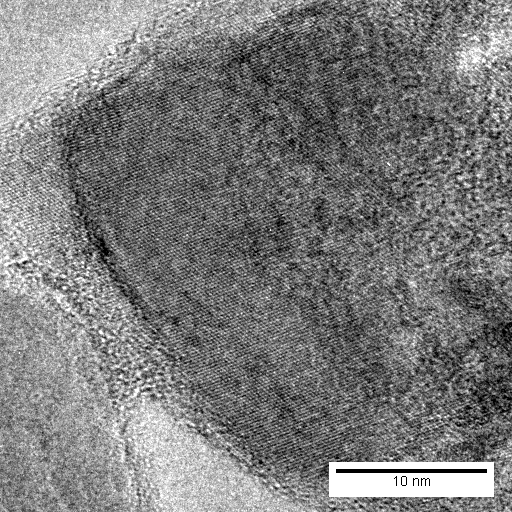}}
					\caption{}\label{subfig::tem2}
				\end{subfigure}
				\caption{Morphology of the milled \nds (sample \insituH). (a) SEM graph showing the distribution of the \nd crystals on the substrate. (b,c) TEM graphs of individual diamond nanoparticles, showing the presence of grain boundaries and twinning as well as the typical lattice structure of diamond.}
				\label{fig::semtem}
			\end{figure}

		  	\subsection{Nanodiamond Production} \label{subsec::ndproduction}

				In the following, we describe the processes used to produce \nd samples of interest.
				We deploy two different methods of incorporating \sivs into wet-milled \nds: Incorporation during (\textit{in-situ}), and implanting after the diamond growth process.
				The obtained samples are prefixed \textit{in-situ} and \textit{implanted} respectively, see \cref{tab::samplenames} for an overview of the available samples.

\begin{table}[tp]
				\centering
				\caption{Overview of the investigated \nd samples. The columns indicate sample names, the mean diameter of the \nds, the \siv incorporation method, and the post-processing treatment(s) of the samples.} \label{tab::samplenames}
					\begin{tabular}{llll}
					\toprule
					Sample name & Diameter & Siv incorporation & post-processing \\
					\midrule
					\insituF & \SI{50}{nm} & \textit{in-situ} & \begin{tabular}[c]{@{}l@{}}series of individual samples with\\combinations of annealing and oxidation\end{tabular}\\ \hline
					\insituS & \SI{70}{nm} & \textit{in-situ} & \begin{tabular}[c]{@{}l@{}}series of individual samples with\\combinations of annealing and oxidation\end{tabular}\\ \hline
					\insituSn & \SI{70}{nm} & \textit{in-situ} &  \begin{tabular}[c]{@{}l@{}}no post-processing \\ subset of \insituS \end{tabular}\\ \hline
					\insituSo & \SI{70}{nm} & \textit{in-situ} & \begin{tabular}[c]{@{}l@{}}oxidized in air at \SI{450}{\celsius} \\ subset of \insituS \end{tabular}\\ \hline
					\insituH & \SI{100}{nm} & \textit{in-situ} & \begin{tabular}[c]{@{}l@{}}series of individual samples with\\combinations of annealing and oxidation\end{tabular}\\ \hline
					\insituHao & \SI{100}{nm} & \textit{in-situ} & \begin{tabular}[c]{@{}l@{}}annealed in vacuum at \SI{900}{\celsius}, \\ consecutively oxidized in air at \SI{450}{\celsius} \\ subset of \insituH \end{tabular}\\ \hline
					\implantedTao & \SI{250}{nm} & implanted & \begin{tabular}[c]{@{}l@{}}annealed in vacuum at \SI{900}{\celsius}, \\ consecutively oxidized in air at \SI{450}{\celsius}\end{tabular}\\
					\bottomrule
					\end{tabular}
			\end{table}
				The starting material for the wet-milled \nds was a nanocrystalline diamond film \cite{Williams2006a}, directly grown on a silicon wafer using chemical vapor deposition (CVD).
				A microwave hydrogen plasma containing 1\% methane was used to grow on purified \SI{5}{\nano\meter} nanodiamond seeds (produced by PlasmaChem).
				To induce \textit{in-situ} \siv creation, sacrificial \si pieces are situated in the growth chamber.
				During diamond growth the \si pieces are etched by the plasma and individual \si atoms are incorporated into the diamond lattice.
				Using a wet-milling process, the diamond film after removal of the substrate is milled with steel beads in a vibrational mill.
				The resulting particle suspension is fractionated using centrifugation, leading to diamond particles of a size of about  \SIlist{50; 70; 100}{\nano\meter} in average diameter (\cref{subfig::sem}), as determined with dynamic light scattering. Transmission Electron Microscopy (TEM) graphs of the milled diamond particles show that the \nds are polycrystalline and exhibit typical single-crystal sizes of a few tens of nanometers.
				In \cref{subfig::tem,subfig::tem2} TEM images of a typical \nd are shown. Within the \nd, several sharp lines are visible.
				These lines are edges of crystal boundaries and grain boundaries, introducing strain in the diamond lattice.
				We remark at this point that some studies suggest the possibility that \sivs are created with a higher probability at grain boundaries and morphological defects than within the core of the crystal \cite{Bray2016,Zapol2001}.
				 The high amount of debris from milling beads is removed for the most part by extensive acid treatment and the absence of debris shown by spectroscopic characterisation.
We also explored milling \nds with silicon nitride beads, and found that the choice of bead material did not cause any noticeable spectroscopic difference.
				The aqueous solution containing the \nds is drop-cast onto an \ir film on a \si substrate.
				The \SI{130}{nm} \ir film is grown onto a buffer layer of yttria-stabilized zirconia (YSL), which in turn is grown onto a \si wafer.
				The \ir surface has the advantage that it acts as an antenna enhancing the collection efficiency of fluorescence light \cite{Neu2012a}.
				Prior to drop-casting, the substracte was cleaned using Piranha solution (50\% sulfuric acid H$_2$SO$_4$, 50\% hydrogen peroxide H$_2$O$_2$).
				This enhances surface hydrophilicity, leading to a homogeneous distribution of the diamond particles on the substrate.
				Post-processing treatment is comprised of annealing in vacuum at \SI{900}{\degreeCelsius}, consecutive \ox in air at a temperature of \SI{450}{\degreeCelsius} or a combination thereof.
				The duration for either treatment method was 3-6 hours.
				\\
				As mentioned before, we also investigated \nds with \sivs implanted after completed diamond growth.
				As starting material we used a polycrystalline diamond film (Element Six, electronic grade) for which we verified that it did not contain \sivs initially.
				In bulk material, the implantation causes the \sivs to form in a specific depth dependent on the implantation energy, leaving most of the diamond vacant of \sivs.
				As a consequence, a significant portion of \nds milled from such a bulk material would not host any \sivs.
				To obtain diamond particles with a homogeneous distribution of \sivs, the following steps were taken:
				First, the diamond film was milled to diamond particles of sizes on the order of a few micrometers.
				In a second step, these microdiamonds were densely spin-coated onto \ir substrates and implanted with \si (implantation energy \SI{900}{keV}, fluence \SI{e11}{\per\centi\meter\squared}).
				To eliminate damage from the implantation process, the diamonds were annealed in vacuum at \SI{900}{\degreeCelsius} and subsequently oxidized in air at \SI{450}{\degreeCelsius} for 3 hours each.
				At this stage, we verified successful creation of \sivs via optical spectroscopy.
				Finally, the micrometer sized diamond particles were milled to a size of \SI{250}{\nano\meter}.

			\subsection{Experimental Setup} \label{subsec::setup}

				We obtained \pl spectra and photon statistics of the samples using a home-built confocal microscope.
				For excitation we use a continuous wave diode laser (Sch\"after-Kirchhoff, 58FCM) at \SI{660}{\nano\meter}.
				The excitation laser is focused onto the sample via a microscope objective with a numerical aperture of 0.8 (Olympus, LMPlanFLN 100x), which also collects the fluorescence light emitted from the sample.
				Both the laser light reflected by the sample and the fluorescence light pass through a glass plate used to couple the excitation laser into the microscope objective.
				The residual laser light is then filtered out by two \SI{710}{\nano\meter} longpass filters.
				The emission light is coupled into a single mode fiber which serves as pinhole in the confocal setup.
				The emission is either guided to a grating spectrometer or to two single photon detectors (PicoQuant, tau-SPAD-100) used in a \hbt configuration to measure photon statistics.
				In front of the avalanche photo diodes bandpass filters select the spectral window in which the investigated color centers emit photons.
				These filters are chosen according to each individual emitter:
				Due to strong shifts in the \zpl wavelength, we use different bandpass filters in front of the APDs to suppress \bkg fluorescence and effectively select the luminescence of the \zpl.

	\section{Results}	\label{sec::results}

	In the following, we present our findings regarding diamond crystal characteristics and spectroscopic measurement of \sivs.
	Unless explicitly stated, we rely on measurements of milled \nds containing \textit{in-situ} incorporated \sivs, i.e.\ samples \insituF, \insituS, and \insituH as listed in \cref{tab::samplenames}.

	\subsection{Diamond Crystal Quality}\label{subsec::raman}

			\begin{figure}
				\begin{subfigure}[tp]{0.45\linewidth}
					\centering
					\testbox{\includegraphics[trim = 0 0 0 0 , clip = true, width = \linewidth]{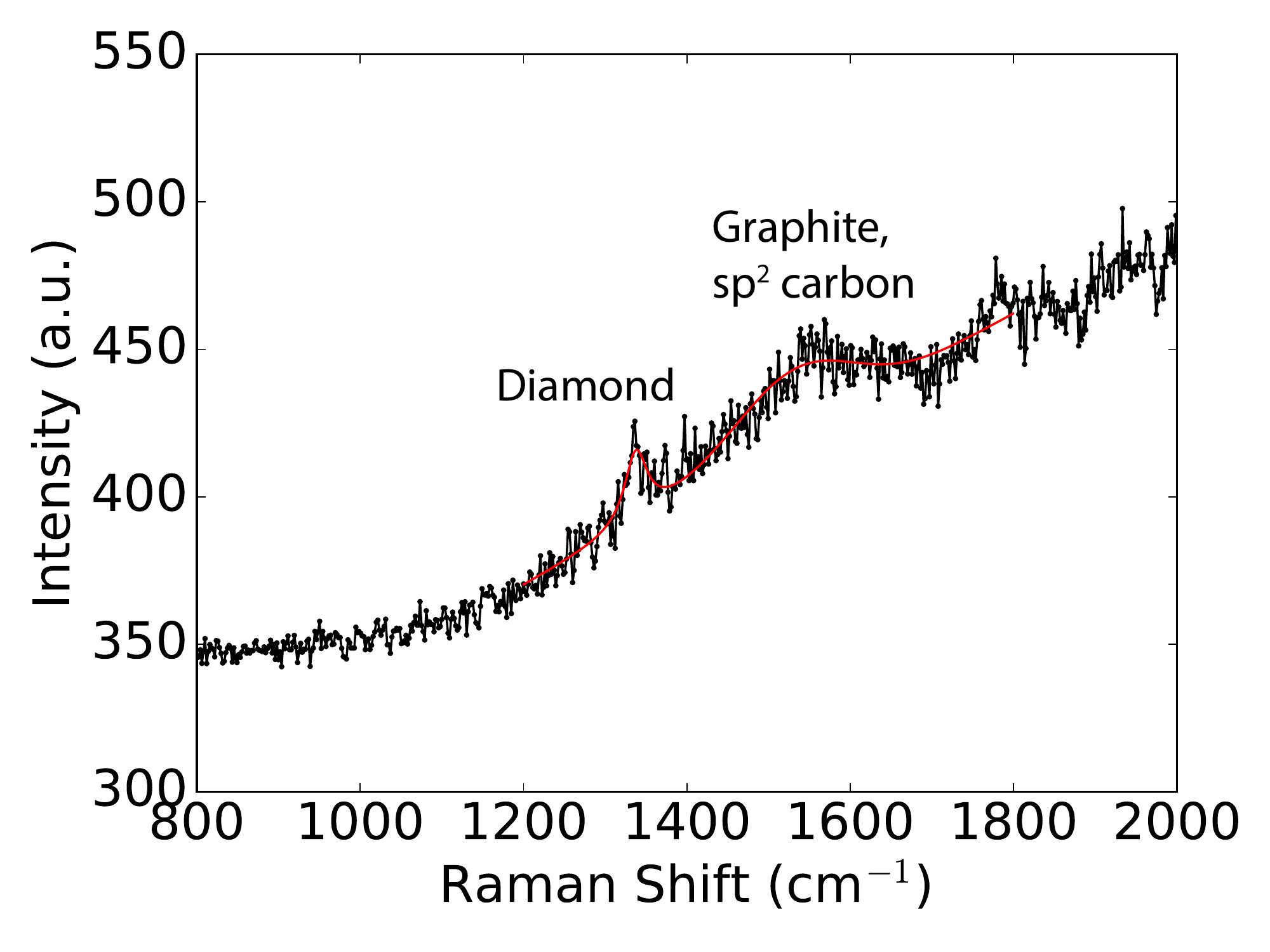}}
					\caption{}\label{subfig::raman_no}
				\end{subfigure}
				\hfill
				\begin{subfigure}[tp]{0.45\linewidth}
					\centering
					\testbox{\includegraphics[trim = 0 0 0 0,  clip = true, width = \linewidth]{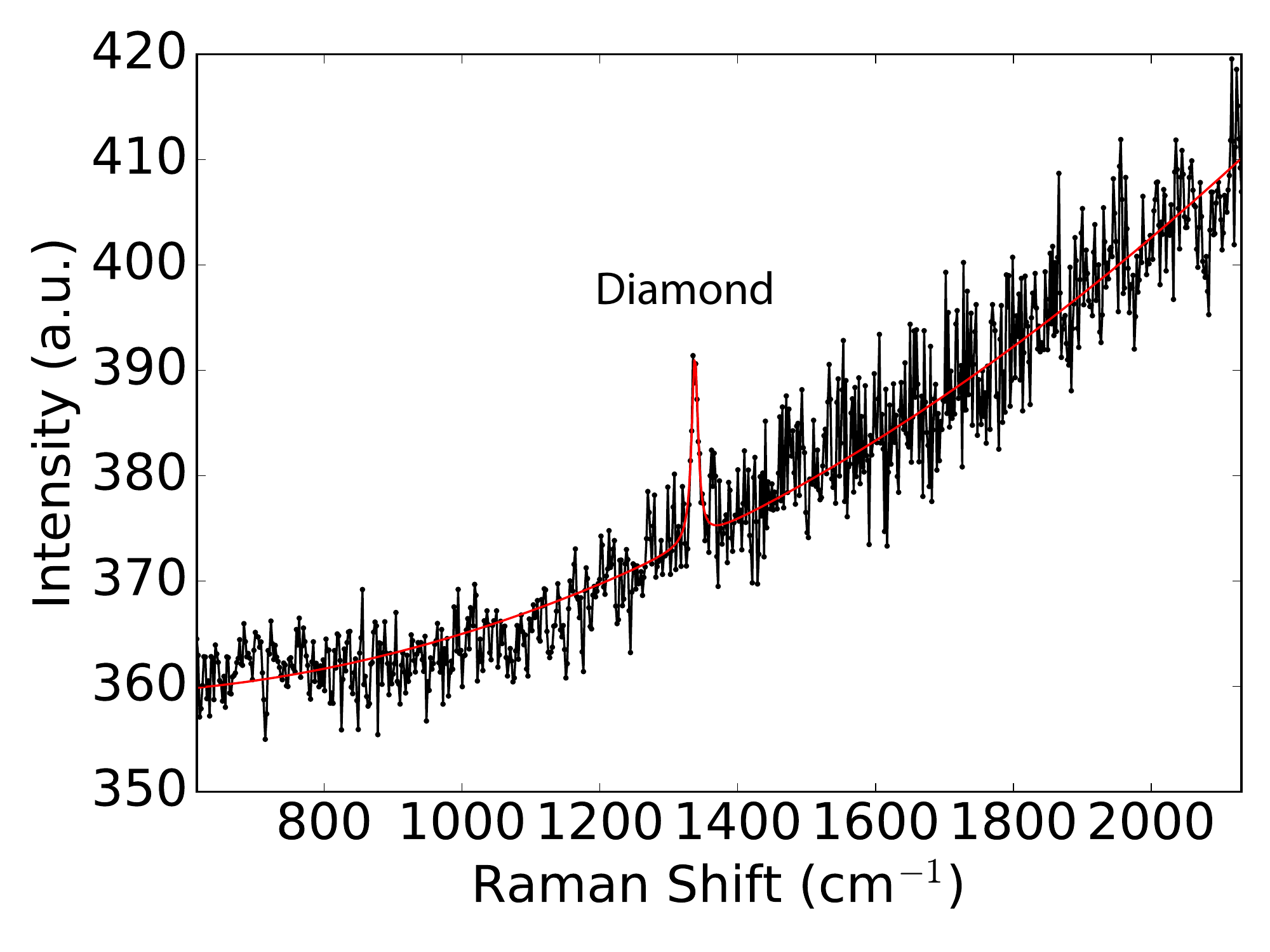}}
					\caption{}\label{subfig::raman_ox}
				\end{subfigure}
                \caption{Raman measurements, black: data, red: fit. (a) Raman measurement before oxidation, sample \insituS. The diamond Raman peak is situated at \SI{1338}{\per\centi\meter}. The broad feature around \SI{1580}{\per\centi\meter} corresponds to the graphite G-band. (b) Raman measurement after oxidation, sample \insituSo. The G-band has vanished, indicating removal of the majority of graphite and amorphous sp$^2$ hybridized carbon.}
				\label{fig::raman}
			\end{figure}

			We aim to perform spectroscopic measurements of single \sivs in wet-milled \nds.
			To this end, we focus on producing pristine wet-milled diamond nanoparticles containing a single \siv each.
			Raman measurements of the \nds allow us to identify issues with surface contamination, defects of the diamond lattice, and strain in the diamond lattice \cite{Zaitsev2001,Prawer2004,Orwa2000}.
			Surface contamination like graphite and amorphous sp$^2$ hybridized carbon manifest themselves as additional peaks in the Raman spectrum.
			Strain in the diamond lattice broadens the first order Raman peak and causes it to shift to higher or smaller wavenumbers.
			Similarly, high concentrations of lattice defects cause additional peaks, a broadening of the first order Raman peak and a shift towards smaller wavenumbers.
			\\
			The size of single \nds is on the order of tens of nanometers, thus low signal intensities can become an issue.
			To overcome this problem we pursue two different approaches to perform Raman measurements:
			\begin{enumerate}[label=\alph*),ref=\alph*)]
				\item \Nd clusters: \label{item::raman_gband} Collective measurements are carried out at several areas on the sample \insituS. Since this sample is densely covered with \nds, collective measurements of clusters of \nds (\cref{subfig::raman_no}) achieves higher signal intensities.
				\item Large \nds: \label{item::raman_implanted} Raman measurements are carried out on the implanted sample \implantedTao. For this sample, diamond particles are large enough to yield sufficient intensities on single \nds.
			\end{enumerate}
			For all Raman measurements a \SI{532}{\nano\meter} continuous wave diode laser was used for excitation.
			\\
			\subsubsection*{Surface Contamination}\label{subsubsection::raman_surface_contamination}
			We test the impact of oxidation treatment as described in \cref{sec::methods} on surface contamination.
			\cref{subfig::raman_no} shows a measured Raman spectrum of a sample without oxidation treatment (\insituSn).
			To verify reproducibility, the measurement is performed on three different spots of the sample.
			The narrow peak in \cref{subfig::raman_no} corresponds to the first order diamond Raman peak and will be further analyzed in below.
			The spectrum also shows a broad peak with a Raman shift of about \SI[separate-uncertainty]{1582+-5}{\per\centi\meter}.
			This shift corresponds to the G-band due to amorphous sp$^2$ hybridized carbon atoms and graphite.
			The exact G-band position and \lw is sensitive to parameters such as the clustering of the sp$^2$ phase, bond-length and bond-angle disorder, presence of sp$^2$ rings or chains, and the sp$^2$/sp$^3$ ratio \cite{Ferrari2004}.
			The \nd Raman spectra are considerably modified after \ox in air at \SI{450}{\degreeCelsius}.
			To verify this, we perform Raman measurements on three different spots of a sample produced in the same process as the above mentioned, which is additionally oxidized (\insituSo).
			While the G-band peak is present in every measurement performed on a non-oxidized sample, it is not present in any of the oxidized samples (\cref{subfig::raman_ox}), indicating successful removal of a majority of sp$^2$ hybridized carbon and surface graphite.
			\\
			\subsubsection*{Defect Concentration}\label{subsubsection::raman_defect_concentration}
			Several effects impact the first order diamond Raman line:
			\begin{enumerate*}
				\item defects in the diamond lattice,
				\item hydrostatic pressure,
				\item uniaxial or more complicated stress configurations.
			\end{enumerate*}
			In the measurement on \nd clusters the width of the diamond Raman peak of sample \insituS varies between \SIlist{15; 30}{\per\centi\meter} without \ox treatment, but is only \SIrange{9}{11}{\per\centi\meter} after the \ox process.
			A possible reason for this change of the width is improved crystal quality \cite{Neu2011a,Prawer2004}.
			In the measurement on large \nds we measured a Raman line at \SI[separate-uncertainty]{1308+-5}{\per\centi\meter} (denoted line R1) which exhibits a broad \lw of \SI[separate-uncertainty]{25+-5}{\per\centi\meter}.
			One plausible explanation for both the position and the \lw of the Raman line are defects in the diamond lattice\cite{Prawer2004}.
			\\
			\subsubsection*{Strain}\label{subsubsection::raman_strain}
			We investigated how strain in the diamond lattice manifests itself in both measurements on \nd clusters and on large \nds.
			In the Raman measurement on \nd clusters, the position of the diamond Raman peak is the same for oxidized (\insituSo) and non-oxidized (\insituSn) samples, indicating that oxidation does not affect strain in the diamond.
			However, the Raman shift of both non-oxidized and oxidized samples amounts to \SI[separate-uncertainty]{1338+-5}{\per\centi\meter}, as compared to the literature value of \SI{1332.5}{\per\centi\meter} of pristine diamond \cite{Zaitsev2001} (given uncertainties are governed by spectrometer resolution).
			This shift indicates the presence of strain in the diamond particles.
			\\
			Performing the Raman measurement on large \nds we find diamond Raman lines between \SI[separate-uncertainty]{1308+-5}{\per\centi\meter} (line R1) and \SI[separate-uncertainty]{1348+-5}{\per\centi\meter} (line R2), indicating a broad distribution of strain among the individual diamond particles (uncertainties governed by spectrometer resolution).
			Only line R1 could be explained with a high defect concentration in the diamond lattice due to its shift to smaller wavelengths.
			However, a more consistent model which explains all occurring shifts is the presence of strain/stress in the diamond nanoparticles.
			The Raman shift $\Delta \tilde{\nu}$ in the presence of compressive and tensile stress is given by \cite{Widmann2016,Liscia2013}:	$\Delta \tilde{\nu} = p/0.34$,		
			where the Raman shift $\Delta \tilde{\nu}$ is given in cm$^{-1}$ and the stress $p$ in GPa.
			The calculation yields a pressure range from \SI[separate-uncertainty]{-8.33+-1.7}{\giga\pascal} tensile stress to \SIlist[separate-uncertainty]{5.27+-1.7}{\giga\pascal} compressive stress.
			Whereas under hydrostatic pressure the triply degenerate first order Raman peak remains degenerate,  under uniaxial and more complex stress configurations (biaxial stress, shear stress etc.) mode splitting occurs \cite{Prawer2004}.
			As mentioned above, we observe broad \lws up to \SI[separate-uncertainty]{25+-5}{\per\centi\meter}.
			The broad Raman \lws may be attributed to uniaxial strain where mode splitting manifests itself in a broadening of the peak due to limited spectrometer resolution.
			\\

		\subsection{Photoluminescence spectra} \label{subsec::spectra}

		To identify \nds containing \sivs, we performed confocal scans of the samples.
		To reduce bias in the measurements, not only the brightest spots of the confocal scans are investigated, but also those which barely exceed \bkg fluorescence.
		\sivs are further investigated by measuring \pl (PL) spectra, single photon statistics and photostability.
		The typical luminescence spectrum of an \siv is composed of a prominent \zpl and weak sidebands.
		Investigations of both are reported independently in the following paragraphs.

		\subsubsection{\Zpl}\label{subsubsec::zpl}

			\begin{figure}[tp]
				\begin{subfigure}[b]{ 0.7\linewidth}
					\centering
					\testbox{\includegraphics[trim = 0 0 0 0,  clip= true, width = \textwidth]{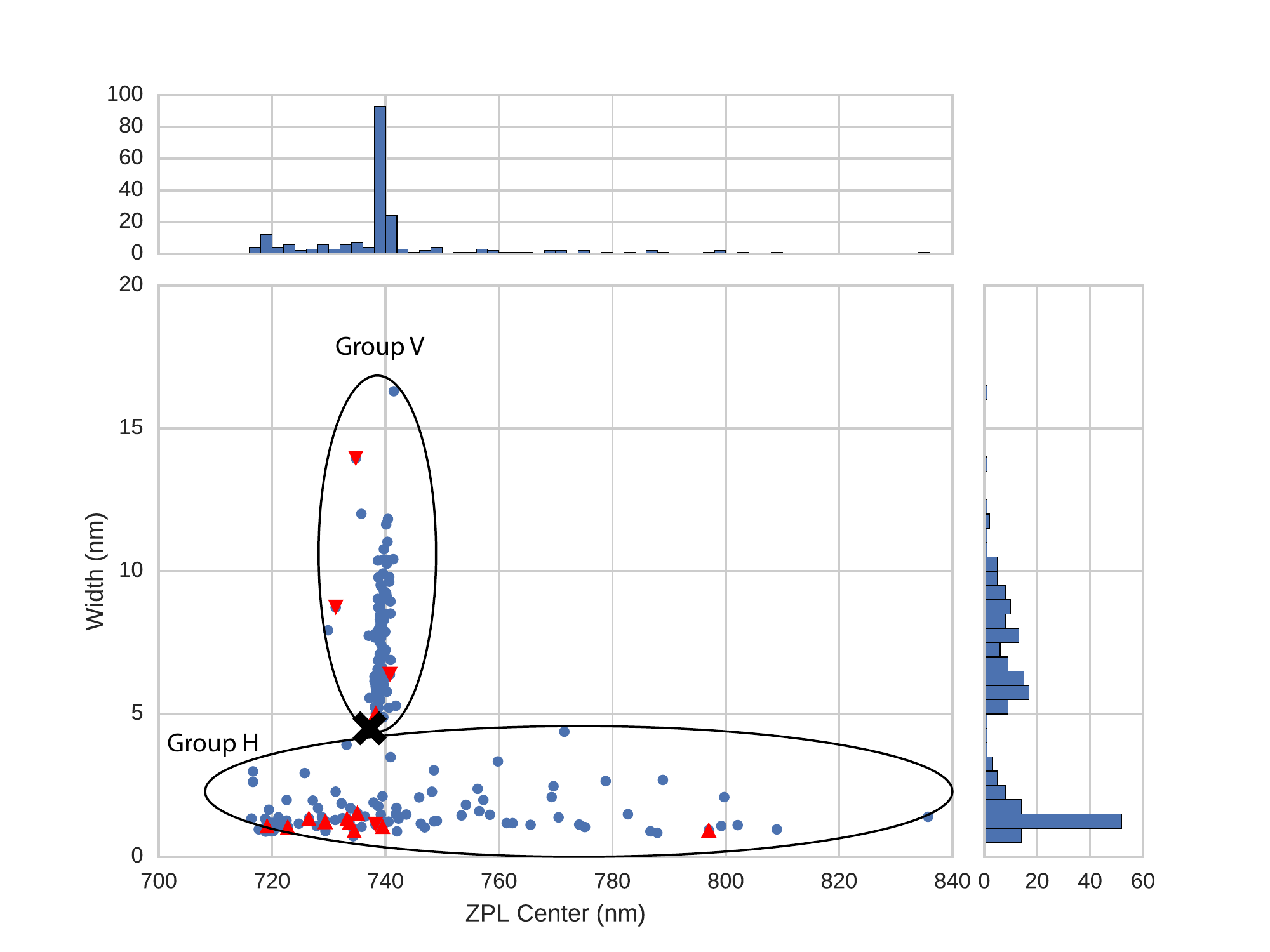}}
					\caption{}
					\label{subfig::distro_big}
				\end{subfigure}
				\hfill
				\begin{subfigure}[b]{ 0.3\linewidth}
					\centering
					\testbox{\includegraphics[trim = 0 0 0 0,  clip= true, width=1\linewidth]{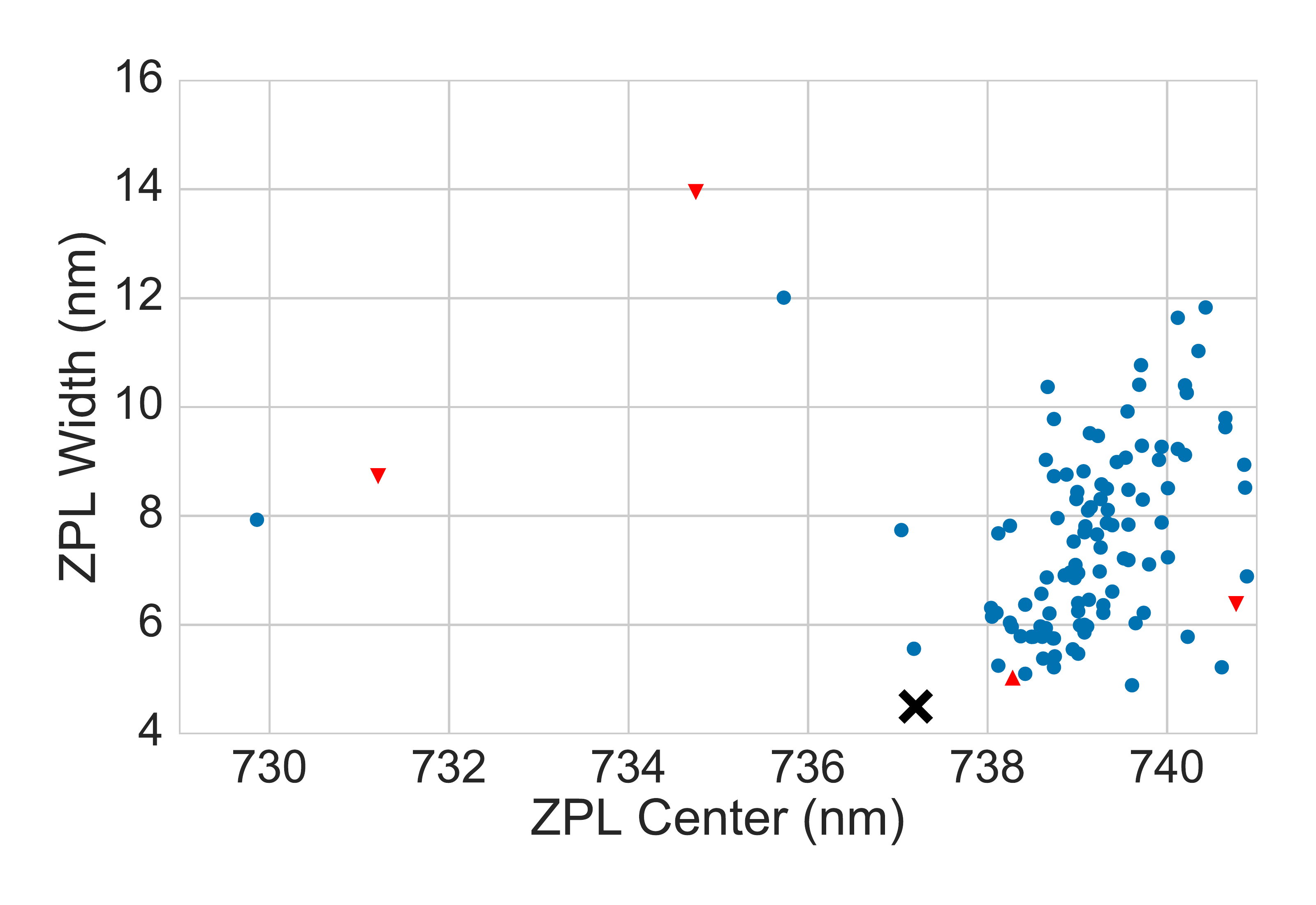}}
					\caption{}
					\label{subfig::distro_inset1}

					\vspace{2ex}

					\centering
					\testbox{\includegraphics[trim = 0 0 0 0,  clip= true, width=1\linewidth]{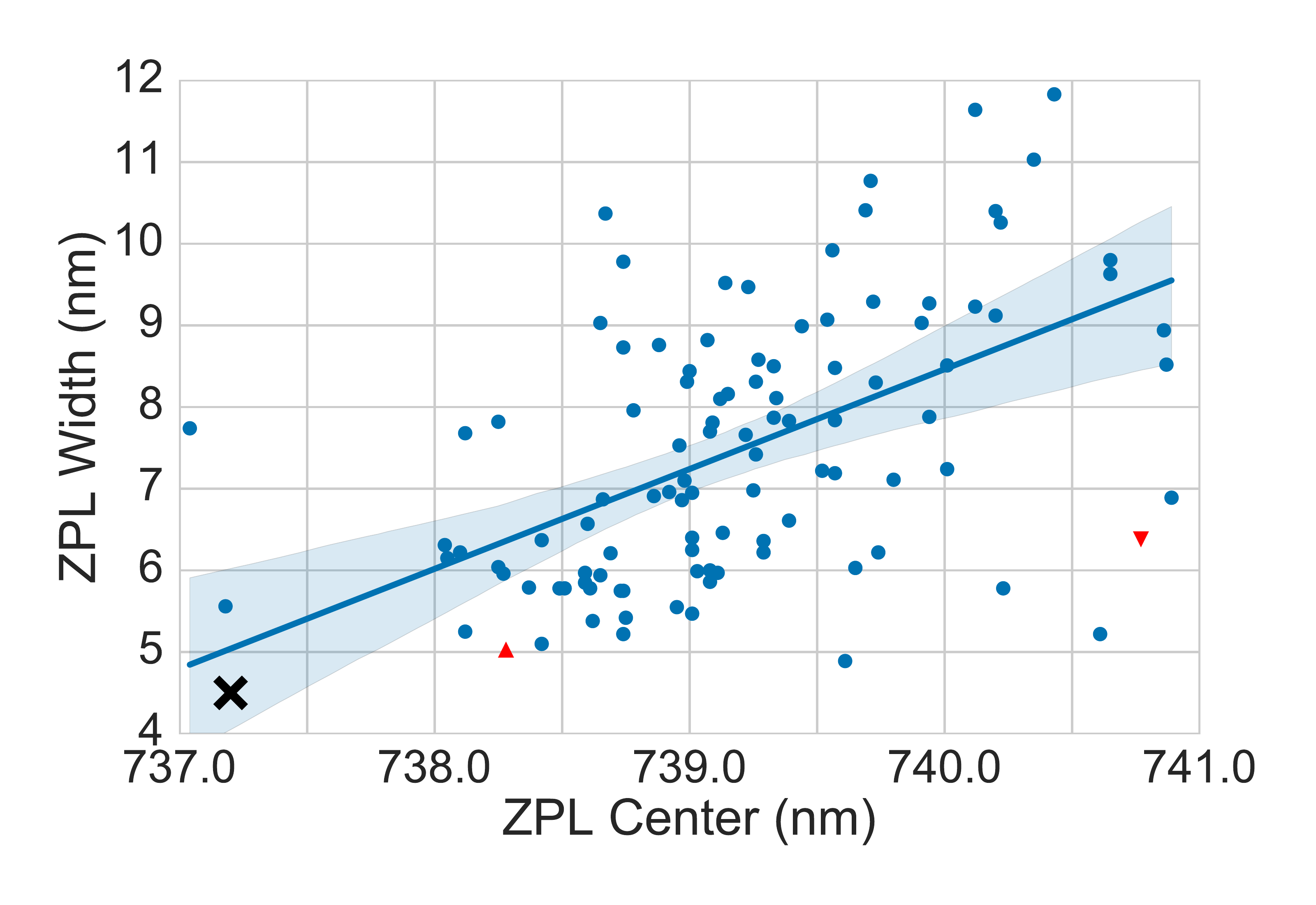}}
					\caption{}
					\label{subfig::distro_inset2}
				\end{subfigure}
				\caption{(a) Distribution of the \ZPL \lw versus the \cwl of the investigated \sivs in milled \nds containing \textit{in-situ} incorporated \sivs for samples \insituF, \insituS, \insituH{}. The data seperates into a horizontal (\hl) and a vertical (\vl) cluster. The bold black cross marks the position of an ideal \siv in unstrained bulk diamond \cite{Arend2016a}. The red triangles indicate emitters with an antibunching dip in the \gtz measurement. Upwards pointing triangles represent blinking emitters (fluorescence intermittency), while triangles pointing down represent non-blinking emitters (see \cref{subsec::photostab}). (b) A zoom into \vl. While many data points exhibit higher \cwls (i.e. a redshift) than the ideal \siv in bulk, only few exhibit shorter \cwls (i.e. a blueshift). (c) Zooming further into \vl, a clear trend of broader \ZPL \lws for larger \ZPL center shifts is visible. The line is a linear regression to all datapoints between \SIlist{737;741}{\nano\meter} which exhibit a \lw larger than \SI{4}{\nano\meter}.}
				\label{fig::bimodal_distr}
			\end{figure}

			\begin{figure}[tp]
				\begin{subfigure}[tp]{0.45\linewidth}
					\centering
					\testbox{\includegraphics[trim = 0 0 0 0 , clip = true, width = \linewidth]{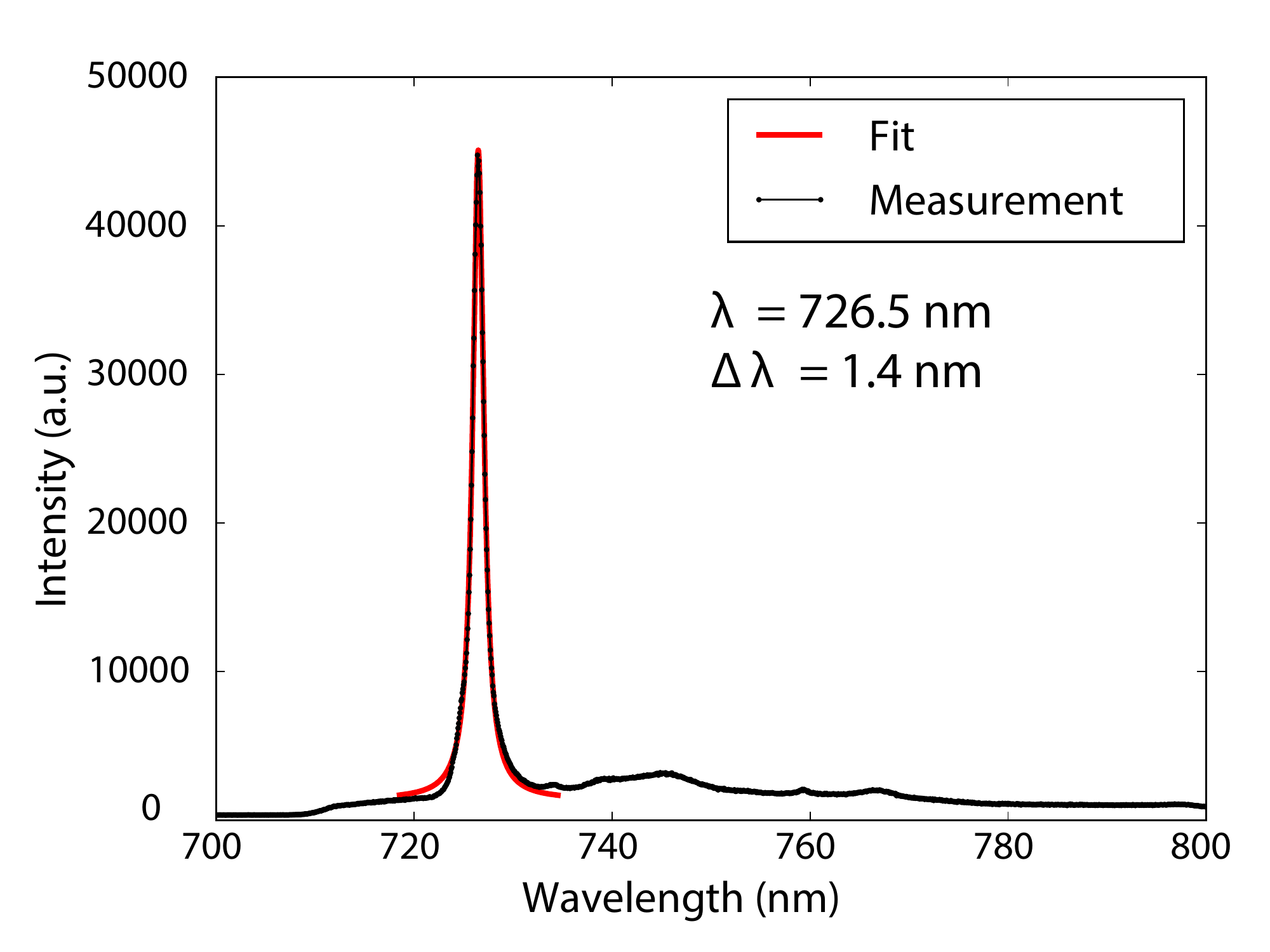}}
					\caption{}\label{subfig::emnarrow}
				\end{subfigure}
				\hfill
				\begin{subfigure}[tp]{0.45\linewidth}
					\centering
					\testbox{\includegraphics[trim = 0 0 0 0,  clip = true, width = \linewidth]{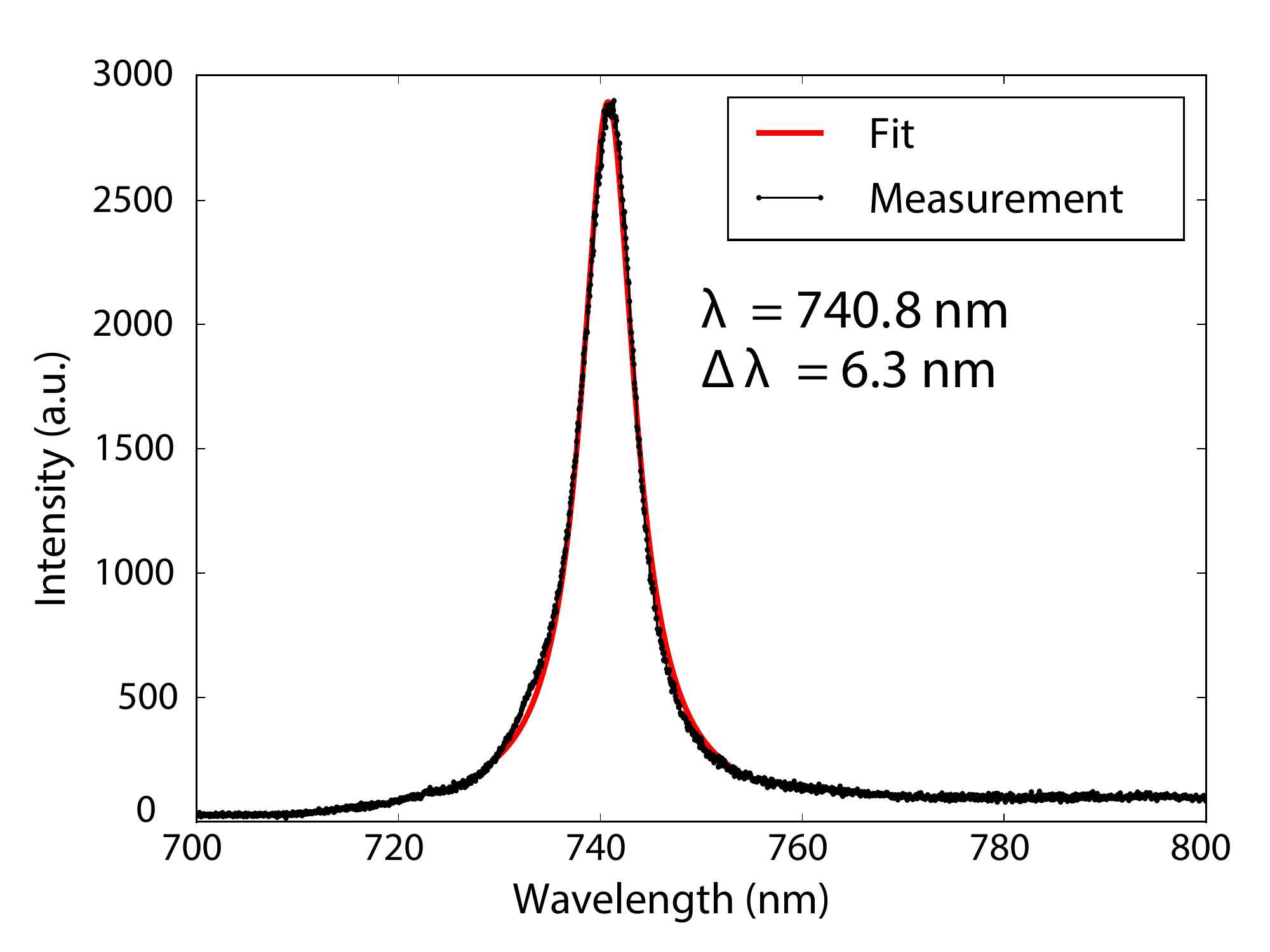}}
					\caption{}\label{subfig::embroad}
				\end{subfigure}
				\caption{Representative photoluminescence spectra of sample \insituHao measured at room temperature. (a) Spectrum of \hl of \cref{fig::bimodal_distr}, denoted \emnarrow. (b) Spectrum of \vl of \cref{fig::bimodal_distr}, dentoted \embroad. The red lines are Lorentzian fits to the peaks.}
				\label{fig::spectra}
			\end{figure}

\begin{figure}[tp]
				\centering
				\testbox{\includegraphics[trim = 0 0 0 0,  clip= true, width = 0.7\textwidth]{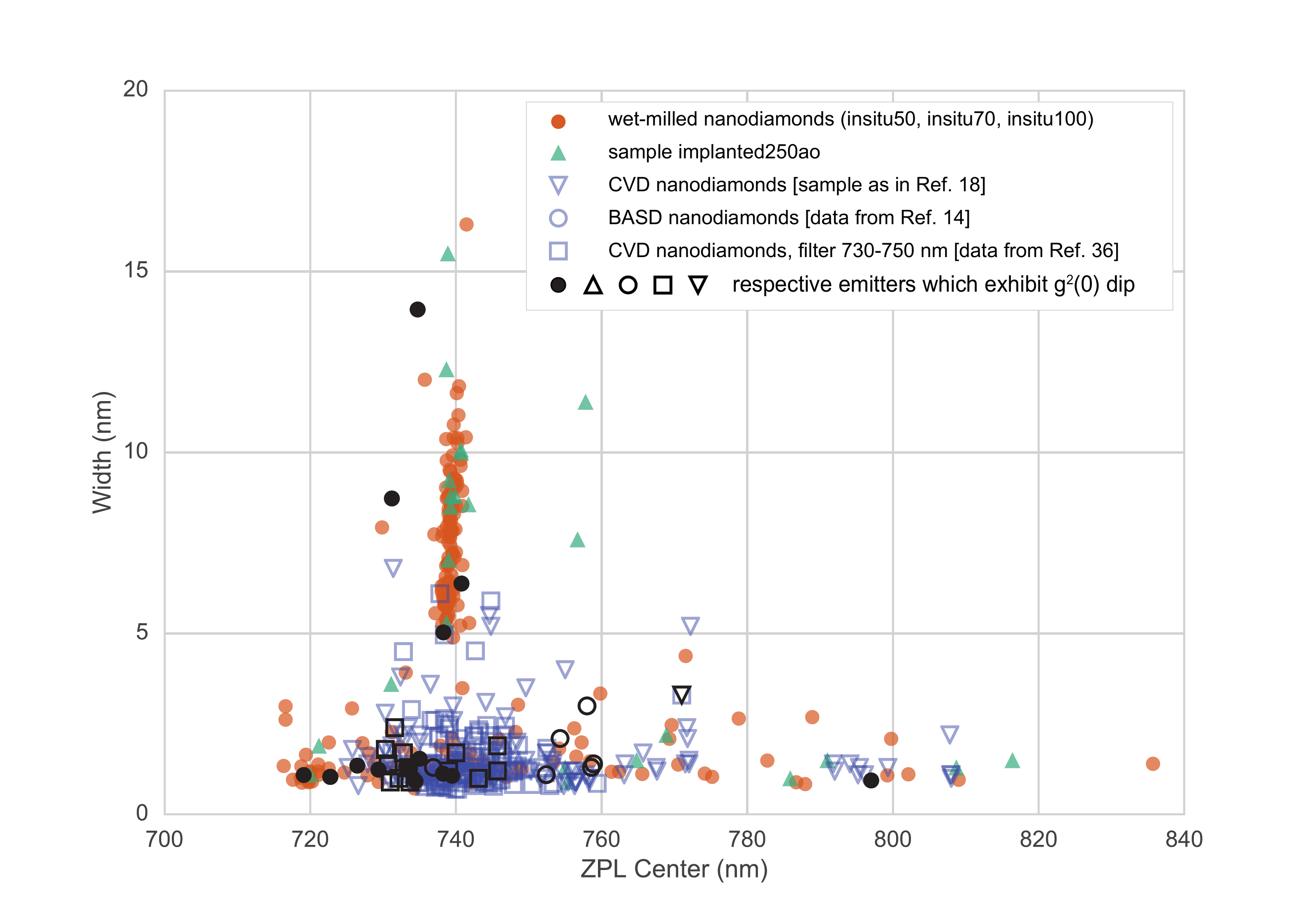}}
				\caption{Comparison of the distribution of the \lw vs. the center wavelength of the ZPL of the investigated \sivs in milled \nds (samples \insituF, \insituS, \insituH) with data measured on: (a) sample \implantedTao (implanted with \si); (b) \nds produced by heteroepitaxial CVD growth, data taken from \cite{Neu2011b} and \cite{Neu2012} where the latter were recorded in a fliter window 730-750 nm only; (c) \nds produced by bead-assisted sonic disintegration of polycrystalline CVD diamond films, data taken from \cite{Benedikter2017a}. Black symbols represent emitters exhibiting a dip in the \gtz function, indicating a single or very few \sivs}
				\label{fig::bimodal_distr_compare}
			\end{figure}

			\paragraph{\ZPL distribution}The \cwl and the \lw of the \zpl (\ZPL) of SiV luminescence spectra for samples \insituF, \insituS, and \insituH are determined by fitting a Lorentzian fit to the \ZPL.
			Both spectra from single and multiple \sivs are taken into account.
			In \cref{fig::bimodal_distr} the \lw for each measured \ZPL is plotted against its \cwl.
			What immediately strikes the eye is a pattern that to our knowledge has not been reported to date:
			The observed \ZPLs partition into two groups, here denoted as horizontal cluster (\hl) and a vertical cluster (\vl). The two distributions are seperated by a gap, i.e.\ a region with a pronounced lack of data points.
			Single emitters are found both in \hl and \vl, marked as red triangles in \cref{fig::bimodal_distr}. Further details on single emitters are given in section \ref{subsec::g2}.
			\\
			The two groups are defined by their characteristic \cwls and \lws:
			In \hl very prominent \ZPL peaks are found showing \lws in the range of \SIrange{1}{5}{nm} and \cwls in the range of \SIrange{715}{835}{nm}.
			\cref{subfig::emnarrow} shows a representative spectrum of a single emitter in \hl (denoted \emnarrow), exhibiting a \ZPL linewidth of \SI{1.4}{nm} and a \cwl of \SI{726.5}{nm}.
			In contrast, in \vl the spectra exhibit broader \ZPL \lws of approximately \SI{5}{nm} up to \SI{18}{nm}.
			Their \ZPL \cwls, however, are distributed within the very narrow range of \SIrange{738}{741}{nm}.
			\cref{subfig::embroad} shows a spectrum of a single emitter of \vl (denoted \embroad) with a ZPL \lw of \SI{6.4}{nm} and a \cwl of \SI{740.8}{nm}.
			For comparison, the room temperature ZPL of \sivs in unstrained bulk diamond exhibits a \lw of \SIrange{4}{5}{nm} and a \cwl of \SI{737.2}{nm} marked with a black cross in \cref{fig::bimodal_distr} \cite{Arend2016a,Dietrich2014}.
			\\
			\paragraph{\db factor}To determine how much the \ZPLs contribute to the total observed emission of \emnarrow and \embroad, we determine the \db factor defined as $DW = I_{ZPL}/I_{TOT}$ where $I_{ZPL}$ and $I_{TOT}$ are intensities in \ZPL and total spectrum, respectively.
			The \db factor for \emnarrow amounts to \num[separate-uncertainty]{0.81(1)} (given uncertainty due to fit).
			This \db factor corresponds to a \hr factor $S =- \ln{(DW)}$ \cite{Walker1979} of \num[separate-uncertainty]{0.21(1)}, which is in good agreement with the values reported in \cite{Neu2011b}.
			The error is mainly due to background corrections.
			When zooming in onto the spectrum of \embroad we do not find distinct sidebands peaks, i.e.\ almost all emission for this emitter is contained within the \ZPL.
			Considering resolution limits of the spectrometer, darkcounts and fluorescence background, we evaluate the \db factor to be larger than \num[separate-uncertainty]{0.97} which is the largest \db factor among all investigated \sivs.
			The two  \db factors discussed here should not be interpreted as  representative for the respective groups, they rather demonstrate the spread of the \db factors of both groups.
			It has to be pointed out that we did not find any systematic difference of the \db factor between \hl and \vl.
			\\
			\paragraph{Comparison to earlier results}To provide context for the novel findings presented in \cref{fig::bimodal_distr}, we compare our results to various earlier findings.
			Furthermore, we discuss an additional comparison to an investigated control sample fabricated using \si implantation.
			The results are presented in \cref{fig::bimodal_distr_compare}.

			Samples for which previous data has been taken are:
			\begin{enumerate}
				\item \nds produced by \basd (BASD) of polycrystalline \CVD diamond films \cite{Neu2011a} (open blue circles in \cref{fig::bimodal_distr_compare}; data taken from \cite{Benedikter2017a})
				\item \label{item::elke_cvd}\nds produced by heteroepitaxial \CVD growth on Ir substrates \cite{Neu2011b} with \textit{in-situ} incorporated \sivs; measured in a spectral filter window of \SIrange{730}{750}{nm} (blue squares in \cref{fig::bimodal_distr_compare}; data reused from \cite{Neu2012} with permission)
				\item \nds as in \ref{item::elke_cvd} (blue downwards pointing triangles in \cref{fig::bimodal_distr_compare}; spectroscopic measurement performed with setup described in \cref{sec::methods}, no filter window)
			\end{enumerate}
			All previous data from different \nd material fit nicely with the \ZPL distribution presented in \cref{subfig::distro_big}, confirming the findings of \cref{fig::bimodal_distr}.
			We verify that the observed luminescent centers are indeed silicon-related defects by performing control experiments with \si implanted samples (sample \implantedTao).
			By doing so we rule out the possibility that the two clusters in the distribution are a result of artifacts.
			Such artifacts include other elements incorporated into the \nds during the growth process: Residue from previous processes performed in the diamond growth chamber or material from chamber parts may be incorporated during \nd growth.
			\cref{fig::bimodal_distr_compare} shows that the implanted \sivs cover the same spectral range from around \SIrange{720}{820}{nm} as the \textit{in-situ} incorporated centers.
			This correlation provides strong evidence for the \si related origin of the defects.
			\\
			\paragraph{Discussion of \ZPL strain shift}To provide a theoretical interpretation, the \ZPL \cwl shift is investigated in further detail and compared to results from density functional theory calculations:
			Zooming in to \vl (\cref{subfig::distro_inset1}) it becomes clear that only six of the measured data points in \vl are situated at a shorter \cwl than the point attributed to an ideal \siv in unstrained bulk material, the shortest wavelength \ZPL being situated at \SI{729.9}{nm}. At the same time, much more data exhibit a \cwl red-shifted to the ideal \siv. This asymmetry suggests that a red-shift of the \ZPL of an \siv is significantly more likely than a blueshift.
					Several mechanisms contribute to the \cwl shift, predominantly hydrostatic and uniaxial strain.
			As discussed in \cref{subsec::raman}, we estimate the stress in our nanodiamond sample from Raman measurements to be on the order of  \SIrange{-8.33}{5.27}{\giga\pascal}.
            In the following, we first discuss the stress/strain shift mechanisms for the \siv before we compare theoretically derived strain shift coefficients to the mentioned range of \ZPL shifts.

\paragraph{Computational method}
To gain insight into the strain mechanism for the SiV centers in diamond, we perform \emph{ab initio} Kohn-Sham density functional theory (DFT) calculations on the strain ZPL shift coupling parameters.
The unstrained model of the negatively charged silicon vacancy defect (SiV$^-$) in bulk diamond is constructed starting from a 512 atom pristine diamond simple cubic supercell within the $\Gamma$ point approximation.
The $\Gamma$ point sampling of the Brillouin zone has proven to be adequate for defects in 512-atom supercells \cite{deak14,kaviani14} providing a sufficiently converged charge density.
The SiV$^-$ defect has S=$\frac{1}{2}$ and it is found to have $D_{3d}$ symmetry with an axis oriented along $\langle111\rangle$.
Standard projector augmented-wave (PAW) formalism together with plane waves are applied, as implemented in the Vienna Ab-initio Simulation Package (VASP) code \cite{kresse93,kresse96,kresse96prb,kresse99paw}.
The geometry optimization is carried out within the constructed supercell by using the Perdew-Burke-Ernzerhof (PBE) \cite{perdew96} DFT functional.
A $420$ eV cutoff is applied for the wave function expansion and a $1260$ eV cutoff for the charge density.  The geometry of the defect is optimized until the forces were lower than $10^{-6}$ eV/\AA$^{-1}$. The $D_{3d}$ symmetry is preserved for both the ground state and the excited state after relaxation.

The ground state of the defect is found to have electronic configuration $e_u^4e_g^3$ (${}^2E_g$) while the excited state is modeled by promoting one electron from the $e_u$ to the $e_g$ level and presents electronic configuration $e_u^3e_g^4$ (${}^2E_u$). Both these states are dynamic Jahn-Teller systems \cite{hepp14, rogers14}. The optical signal of the defect (ZPL) can be calculated as the lowest excitation energy by the constraint DFT approach (CDFT) \cite{gali09}.
According to CDFT one electron is promoted from the ground state to a higher level leaving a hole behind. The interaction between the electron and the hole is included in the procedure. The ZPL energies were obtained by taking the total energies of the optimized geometries in the ground and excited state.

The strain on the defect structure is simulated by applying a compression to the supercell along a well defined direction. The strained supercells are obtained by compressions along $\langle100\rangle$, $\langle110\rangle$ and $\langle111\rangle$. We also study the configuration produced by a hydrostatic pressure, which consists in subjecting the cell to the same compression along the three directions. After introducing the strain along the directions, the ZPL energies were calculated for each strained supercell. Finally, we obtained data points on the calculated ZPL energies vs.\ the applied strain. 
These ZPL energies correspond to the optical transition between the lower branch levels of the ${}^2E_u$ and ${}^2E_g$ doublets. We note that additional calculations were performed for the $\langle100\rangle$ and $\langle110\rangle$ strained supercells by using the screened, range-separated, non-local hybrid density functional of Heyd-Scuseria-Ernzerhof (HSE06) \cite{heyd, krukau} and we found good agreement with the PBE values.

Nudged elastic band (NEB) calculations~\cite{NEB} by HSE06 were performed in order to calculate the energy barriers in the ground state in strained supercells. The barrier energies between the $C_{2h}$ configurations stayed under 10.0~meV which implies small change in the adiabatic potential energy surface around the $D_{3d}$ symmetry upon the applied strain. As a consequence, the Ham reduction factor in strained SiV center will minutely change with respect to that of unstrained SiV center~\cite{Thiering2018}. This suggests that the observed ZPL shifts upon stress are directly strain related, and the contribution of the change of the effective spin-orbit to the ZPL shifts is minor.

\begin{figure}[tp]
				\centering
				\testbox{\includegraphics[trim = 0 0 0 0,  clip= true, width = 1.0\textwidth]{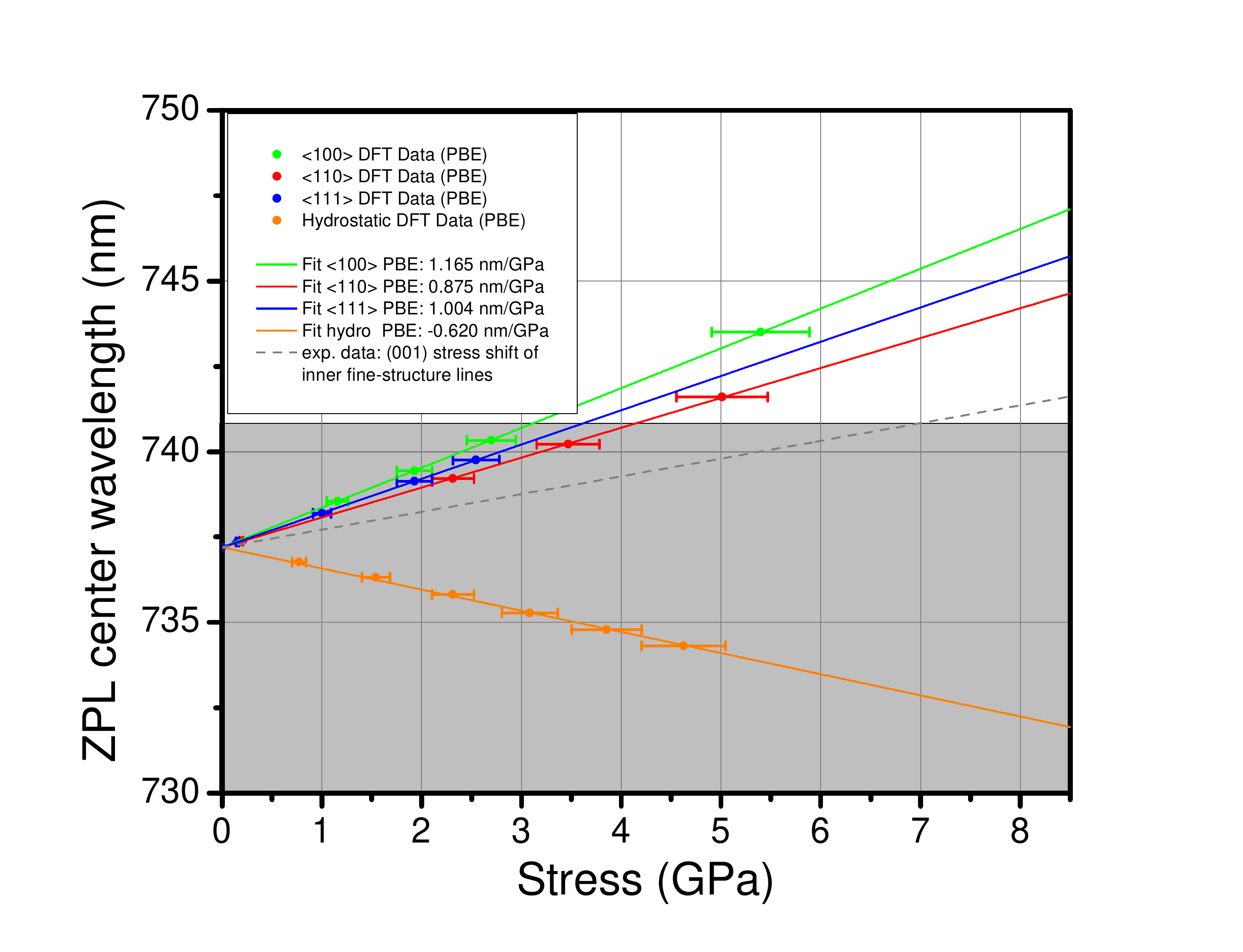}}
				\caption{Calculations of the wavelength of the \siv \ZPL in dependence of pressure. Markers: DFT calculated pressure with PBE and HSE functionals; Lines: linear fits to the calculated points. Hydrostatic-type pressure causes a moderate blue shift whereas uniaxial strain causes larger redshift with different magnitudes depending on the direction of the strain. Grey line: experimental stress shift data for (001) uniaxial stress and inner fine-structure line of the SiV center. Grey area: range of \ZPL \cwls of \vl.}
				\label{fig::stress_pressure}
			\end{figure}

\paragraph{Comparison of theoretical and experimental strain shifts}		The data points in \cref{fig::stress_pressure} show the \ZPL \cwl shifts calculated with the method outlined above. For comparison with experimentally determined stress data the strain values of the theoretical calculation were converted into stress assuming a simplified model where diamond is approximated as isotropic linear elastic material. In this case stress $\sigma$ is related to strain $\varepsilon$ via Young's modulus $E$: $\sigma = E \varepsilon$. This assumption is pragmatic as we do not know the orientation of stress in the nanodiamonds from the Raman measurements but only its modulus. The values of $E$ vary considerably among different diamond materials \cite{Hess2012} but even nanocrystalline diamond may obtain a large Young's modulus $E\geq1000$~GPa \cite{Williams2010}. As an average value for the nanodiamond size used in our investigations we assume $E=1000\pm100$~GPa \cite{Hess2012}. The calculated data points were extrapolated by linear fit functions to yield stress shift coefficients for the range of stress (up to $\approx8.5$~GPa) found in the nanodiamonds. The grey area covers the wavelength range of experimental \ZPL wavelengths measured within group V. The dashed grey line represents stress-shifts of SiV emission lines at low temperatures derived from the only experimentally measured stress shift coefficient for the SiV center under uniaxial stress in $\left<001\right>$ direction \cite{Sternschulte1994,Hepp2014PhD}. The measurements of \cite{Sternschulte1994} were performed at 4~K and reveal the shifts of the SiV center fine structure lines: the outer lines of this fine structure shift with about 4 nm/GPa (2.23 THz/GPa, not shown here), whereas the inner lines shift with only 0.52nm/GPa (292 GHz/GPa, denoted as dashed line). The room temperature spectrum, however, is mostly governed by the inner line "C" of the spectrum \cite{Arend2016a}, i.e. the line with second highest wavelength, corresponding to the optical transition between the lower branch levels of the ${}^2E_u$ and ${}^2E_g$ doublets as used in the DFT calculations. We find that the calculated uniaxial stress shift coefficients match well the experimentally obtained value (dashed line) and both coincide well with the range of measured red-shifted \ZPLs of group V (grey area). We thus interpret the \ZPL shifts of group V as originating from level shifts due to uniaxial strain. Furthermore, the calculated \ZPL shifts due to hydrostatic pressure coincide well with the range of the blue shifted \ZPLs that we observe in group V. The fact that we see only few blue-shifted \ZPLs might be due to reason that pure hydrostatic pressure is rarely observed and overlayed by uniaxial stress in the nanodiamonds.
			On the other hand, the measured shifts in \hl are too broad to be solely explained by strain in the diamond.
			A potential explanation for the very inhomogeneous distribution of defect center \ZPL \cwls could be the association of \sivs with a further nearby defect, such as a vacancy, or a modified SiV complex such as SiV:H \cite{Thiering2015}.
			\\
			Zooming in to \vl, another effect becomes visible (\cref{subfig::distro_inset2}):
			with increasing  \ZPL \cwl, the \lw becomes broader.
			As discussed above, a red-shift of the \ZPL is linked to increasing uniaxial strain.
			Thus we conclude that the \ZPL \lw, too, is affected by strain in the diamond lattice due to a modified electron-phonon coupling \cite{Jahnke2015a}. A similar effect has been previously observed for \sivs at cryogenic temperatures \cite{Arend2016a}.
			\\
			To conclude, we are able to explain the distribution of \ZPL \cwls in \vl very consistently with theoretical predictions based on level shifts due to strain in the diamond lattice.
			On the other hand, we have to assume that \hl is comprised of modified \sivs, the structure of which is currently unclear.

			\subsubsection{Sideband} \label{subsubsec::sideband}

\begin{figure}[tp]
					\centering
					\testbox{\includegraphics[trim = 0 0 0 0,  clip = true, width = 0.49\textwidth]{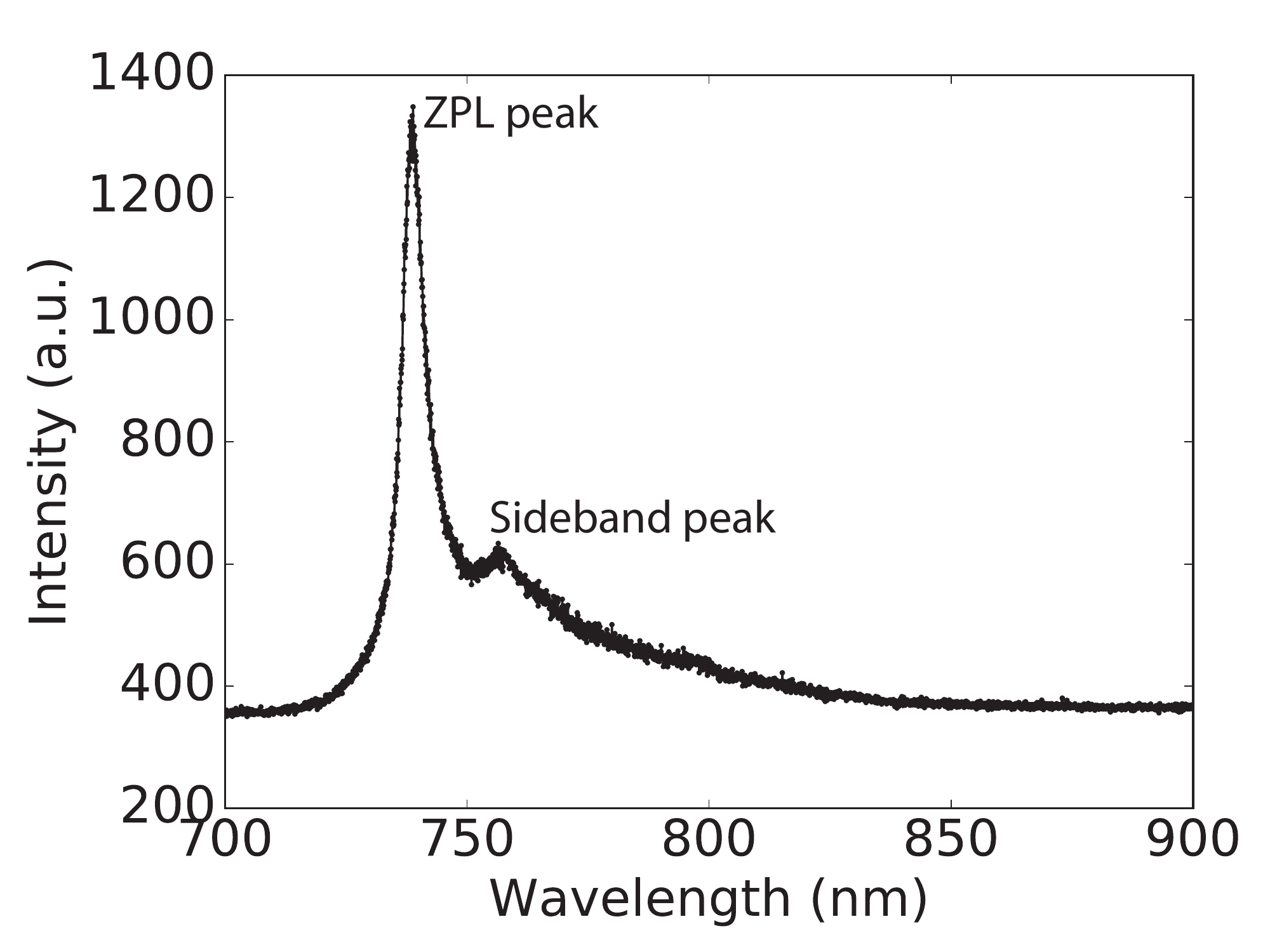}}
					\caption{Representative spectrum of an emitter of \vl exhibiting a sideband peak.}
					\label{fig::broad_peak_sb}
				\end{figure}

				From the literature it is known, that the \siv in \nd exhibits a large \db factor of over \SI{70}{\percent} \cite{Neu2011b,Neu2011}, which is consistent with our measurements of \emnarrow and \embroad.
				Nevertheless, sideband peaks are present in many \siv \pl emission spectra.
				The investigated emitters exhibit two different structures of sideband spectra: The spectra in \vl exhibit one strong sideband peak (\cref{fig::broad_peak_sb}), whereas spectra in \hl exhibit several weaker sideband peaks.
				
\begin{figure}[tp]
					\centering
					\testbox{\includegraphics[trim = 0 0 0 0,  clip= true, width = 0.49\textwidth]{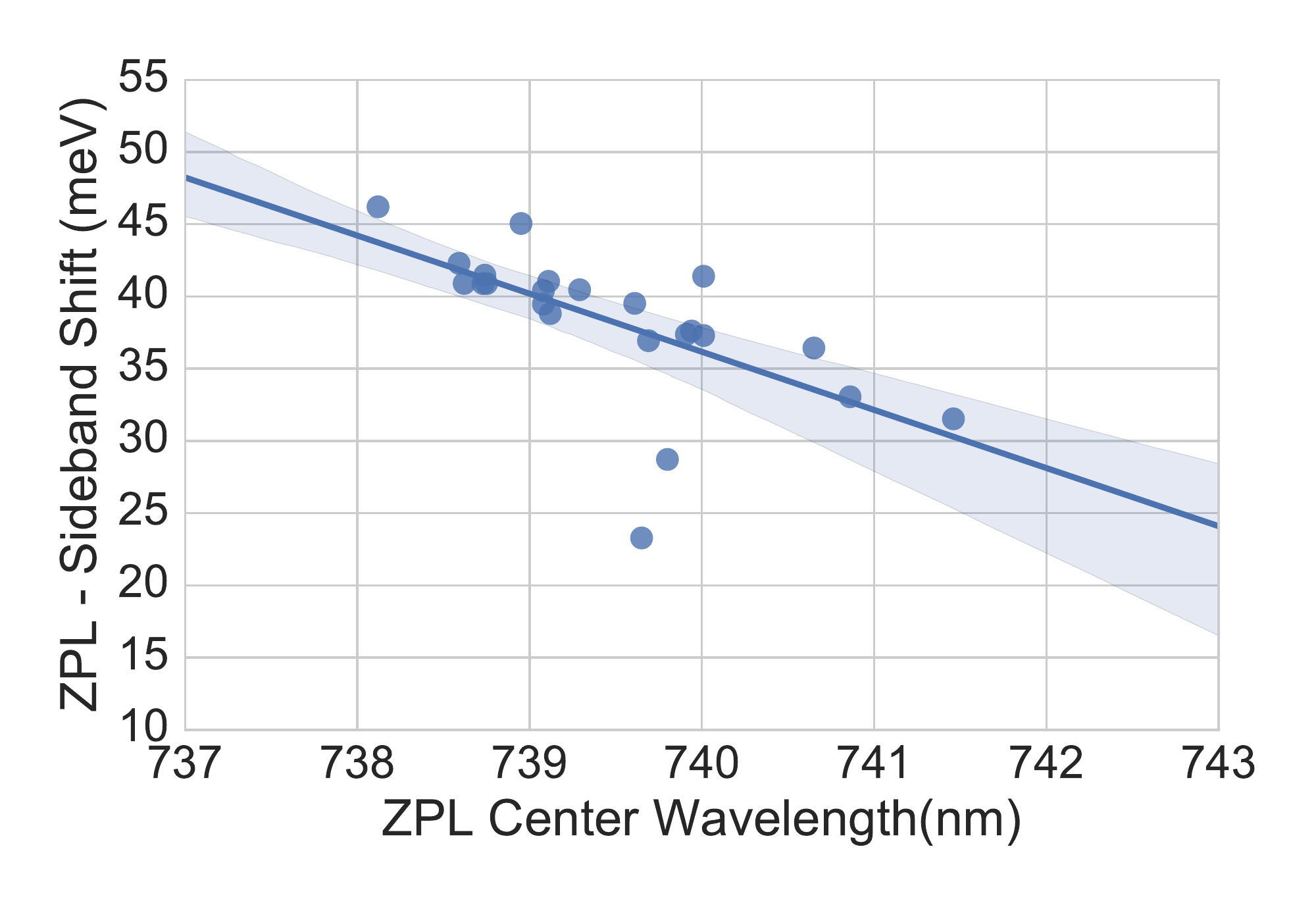}}
					\caption{Shift of dominant sideband peak from the \ZPL in spectra of \sivs (\vl, samples \insituF, \insituS, \insituH) vs. ZPL \cwl. The linear fit shows that the shift decreases with increasing ZPL center wavelength, i.e.\ with increasing strain and exhibits a slope of \SI[separate-uncertainty]{-4\pm1}{\milli\electronvolt\per\nano\meter}. The shaded area is the \SI{95}{\percent} confidence interval.}
					\label{fig::sideband_fit}
				\end{figure}

				About \SI{70}{\percent} of the \vl \pl spectra with one distinct sideband peak display a shift of the sideband peak from the \ZPL between \SIrange{37}{43}{meV}.
				The range of line shifts for the prominent sideband peak coincides with a well-known feature at \SI{42}{meV}, associated with \sivs \cite{Sternschulte1994,Larkins1971}, but also to a larger number of optically active defects \cite{Sternschulte1994}.
				The occurrence of this \SI{42}{meV} sideband feature for a large number of defects and the absence of isotopic variations \cite{Dietrich2014}, favors an assignment as non-localized lattice vibration.
				We furthermore observe that the dominant sideband peak shifts towards smaller distance from the \ZPL for increasing \ZPL \cwl, i.e.\ increasing strain, see \cref{fig::sideband_fit}.

				We fit the data with a linear regression and find a slope of $(-4\pm1)$meV~nm$^{-1}$.
				The low phonon energy of the sideband feature and its shift with strain might arise from a local ``softening'' of the crystal lattice in the vicinity of a defect \cite{Sternschulte1994}.
				A recent study \cite{Londero2016} suggests that the \SI{42}{meV} mode (as other broad sideband features) originates from a resonance of $e_g$.\ phonons causing the dynamical Jahn-Teller effect in the \sivs.
				As the Jahn-Teller coupling varies with strain it is also expected that the resonance shifts accordingly.
				\\
				In the spectra of \vl, we do not observe a typical \siv sideband feature at \SI{64}{meV}, attributed to a local vibration of the \si atom \cite{Dietrich2014}, frequently much stronger than the  \SI{42}{meV} sideband peak.
				A possible explanation is, that the lattice mode at \SIrange{37}{43}{meV} is so strong that the local vibrational mode at \SI{64}{meV} cannot be separated from the tail of the lattice mode.
				\\
				In \hl we observe many spectra which exhibit several peaks within the spectral range of our detection range between \SIrange{710}{900}{nm}.
				The challenge arises to unequivocally distinguish between peaks stemming from a phonon sideband and peaks stemming from shifted, less intense \siv \ZPLs.
								We see, however, a tendency of peaks to accumulate at shifts of around \SIlist{43;64;150;175}{meV}.
				These findings are consistent with sideband shifts reported in \cite{Neu2011,Sternschulte1994,Zaitsev2000}.
				\\

			\subsection{Photon correlation measurements} \label{subsec::g2}

				\begin{figure}[tp]
					\begin{subfigure}[tp]{ 0.49\linewidth}
						\centering
						\testbox{\includegraphics[trim = 0 0 0 0,  clip= true, width = \textwidth]{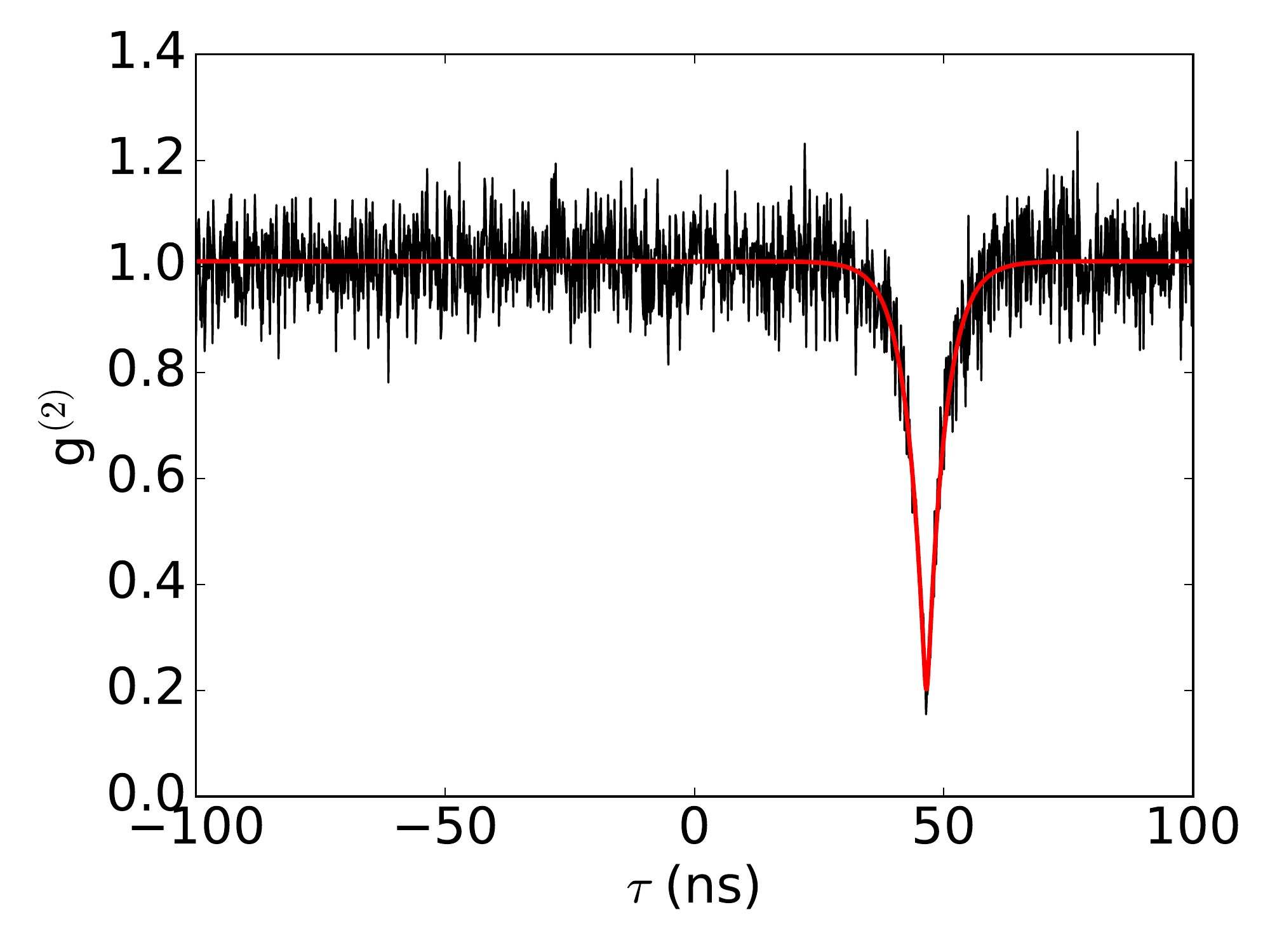}}
						\caption{}\label{subfig::g2_a}
					\end{subfigure}
					\hfill
					\begin{subfigure}[tp]{ 0.49\linewidth}
						\centering
						\testbox{\includegraphics[trim = 0 0 0 0,  clip= true, width = \textwidth]{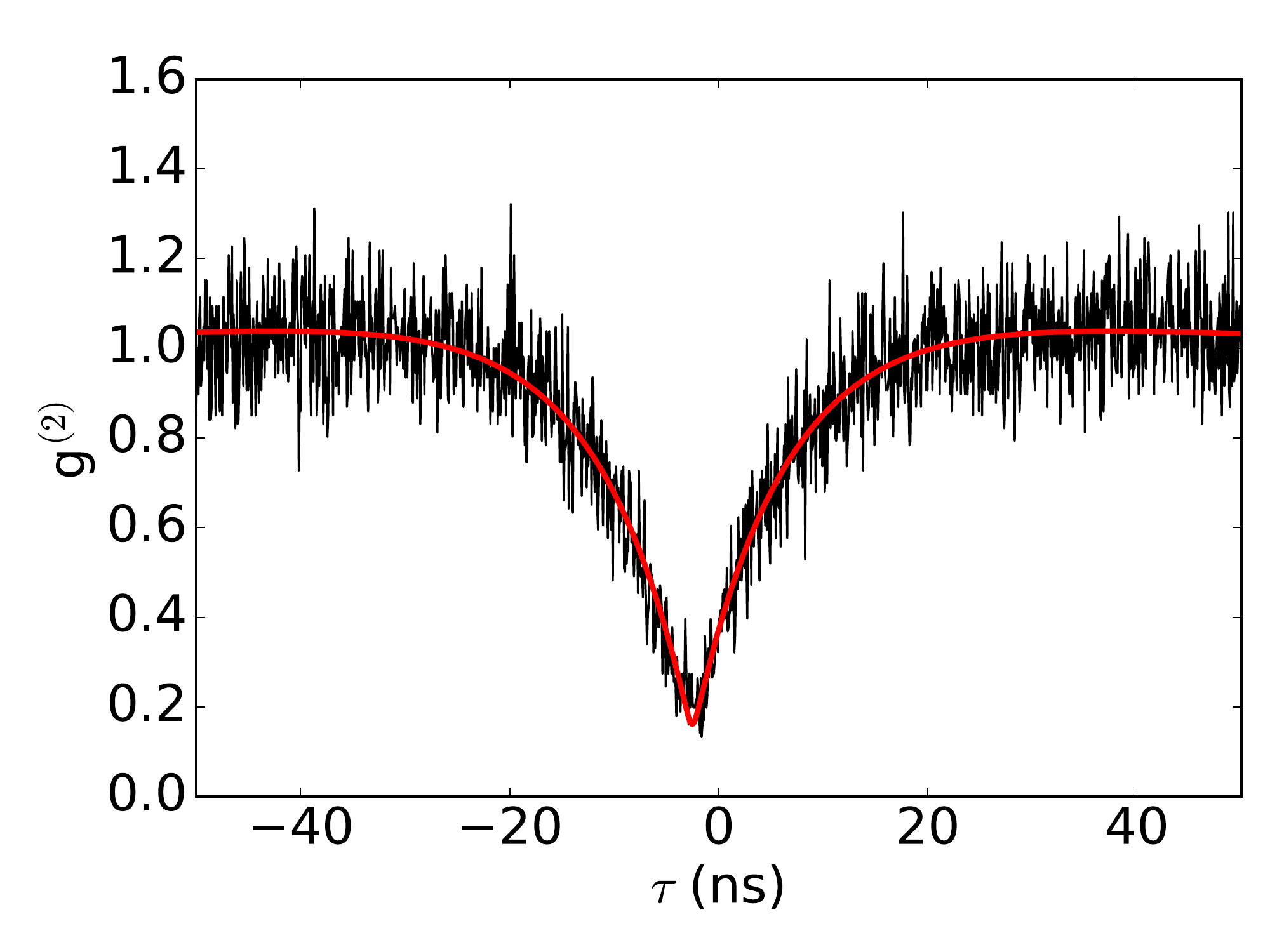}}
						\caption{}\label{subfig::g2_b}
					\end{subfigure}
					\caption{(a) Intensity autocorrelation function of an emitter in \hl. (b) Intensity autocorrelation function of \embroad at an excitation power of \SIlist{200}{\micro\W}(saturation power is \SI{1}{mW}).}
					\label{fig::g2}
				\end{figure}

				The investigated individual \sivs exhibit count rates of a few thousand to a few \num{100000} counts per second (\SI{}{\cps}).
				We carried out measurements of the photon statistics and found that about \SI{3}{\percent} of luminescent nanodiamonds contain single color centers.
				Our measurements further reveal that the probability of finding a single emitter does not correlate in any way with the \cwl or the \lw of the ZPL.
				We found several single \sivs with an antibunching dip down to about \num{0.2} and attribute the residual \gtz value to background fluorescence from the diamond host.
				For the \nds used in our investigations an independent background measurement without simultaneously measuring \siv \pl is infeasible, because the laser spot size is bigger than the \nd.
				Therefore, the background is estimated from the sideband of \siv spectra.
				The measured lifetimes of the single \sivs are in the range of about \SIrange{1}{9}{ns} in accordance with previous research \cite{Sipahigil2014,Neu2011b,Sternschulte1994}.
				\autoref{fig::g2} shows the \gt functions of the two emitters introduced in \autoref{subsec::spectra}, \emnarrow and \embroad.
				\\
				\autoref{subfig::g2_a} shows the photon correlation function of an emitter in \hl.
				The shift of the dip to $\tau=\SI{50}{ns}$ originates from a path length difference of the two detection paths in the \HBT setup.
				The \gtz value of the fit is \num{0.20} due to residual background as discussed above.
				The excited state lifetime of the emitter was determined to be \SI[separate-uncertainty]{3.8\pm0.1}{ns}.
				\\
				\autoref{subfig::g2_b} shows the \gt function of \embroad at an excitation power of \SI{200}{\micro\W}, which is \SI{20}{\percent} of the emitter's saturation power $P_{sat}=\SI[separate-uncertainty]{1.0\pm0.1}{mW}$.
				The \gtz value yields \num{0.16}.
							The lifetime of the excited state of this emitter is \SI[separate-uncertainty]{9.2\pm0.2}{ns} which is the highest excited state lifetime we measured within this work.
				\\
				Several \nd \pl spectra contain multiple narrow distinct peaks at different \wls.
				This circumstance is attributed to \nds containing more than one \siv, each of which is subject to a different \ZPL \wl shift.
				We choose narrow bandpass filters to perform independent measurements of each individual peaks of such a spectrum.
				As a result it is possible to measure \gtz values below \num{0.5} for each of these narrow peaks.
				Hence the individual peaks are identified as single emitters with a different ZPL \cwl.
				\\
				We do not see a systematic difference regarding the photon autocorrelation functions of \hl and \vl, both reach similar \gtz values.
				Also, the timescales of the excited state lifetimes coincide.

			\subsection{Photostability} \label{subsec::photostab}

				\begin{figure}[tp]
					\begin{subfigure}[tp]{ 0.49\linewidth}
						\centering
						\testbox{\includegraphics[trim = 0 0 0 0,  clip= true, width = \textwidth]{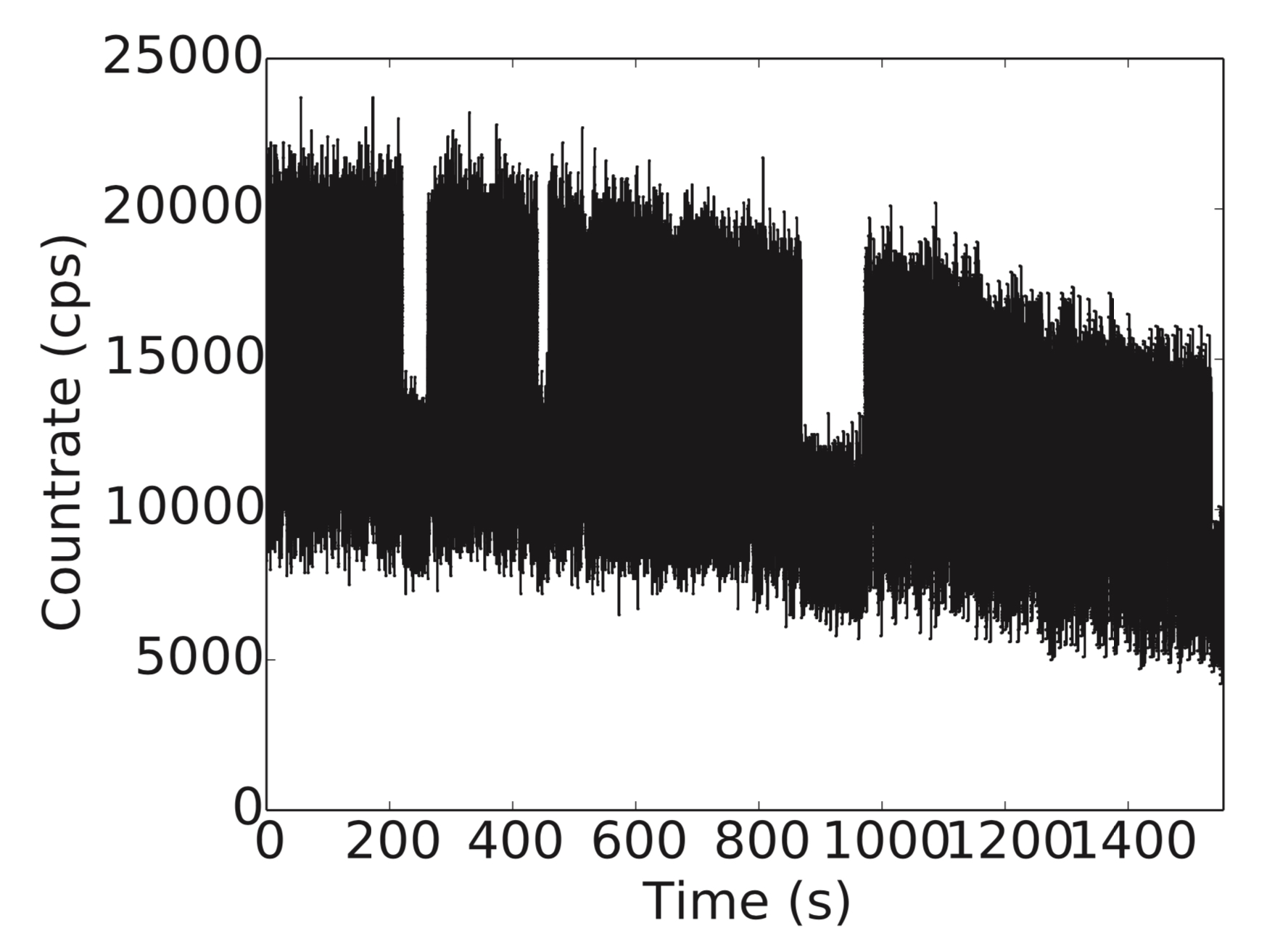}}
						\caption{}\label{subfig::blink_long}
					\end{subfigure}
					\hfill
					\begin{subfigure}[tp]{ 0.49\linewidth}
						\centering
						\testbox{\includegraphics[trim = 0 0 0 0,  clip= true, width =\textwidth]{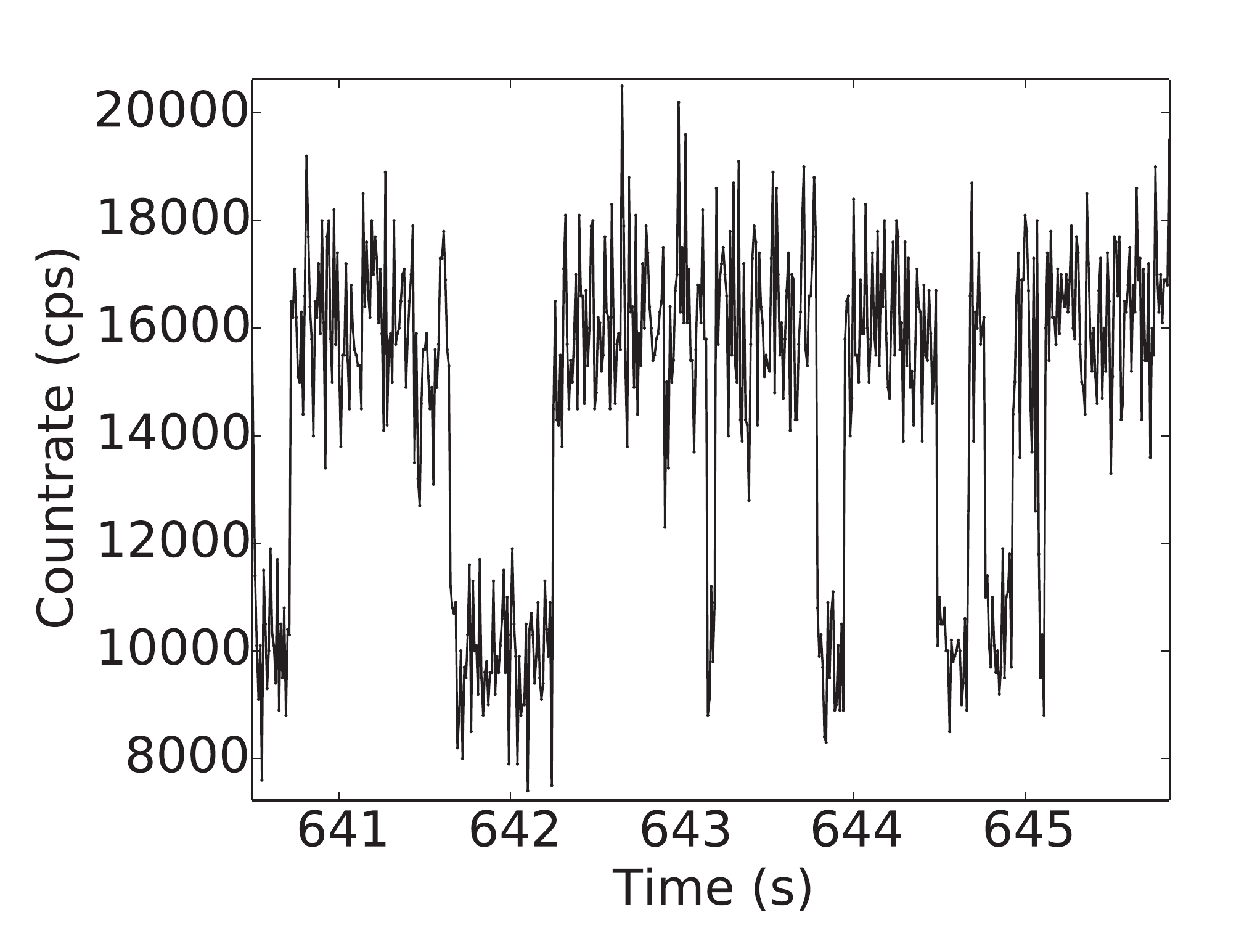}}
						\caption{}\label{subfig::blink_short}
					\end{subfigure}
					\caption{(a) Time trace of the single emitter H1, which exhibits strong blinking. The variation of the count rate in the upper state is attributed to a drift of the setup. (b) Detail of the time trace of the same emitter.}
					\label{fig::blink}
				\end{figure}

				\begin{figure}[tp]
					\centering
					\testbox{\includegraphics[trim = 0 0 0 0,  clip= true, width = 0.7\textwidth]{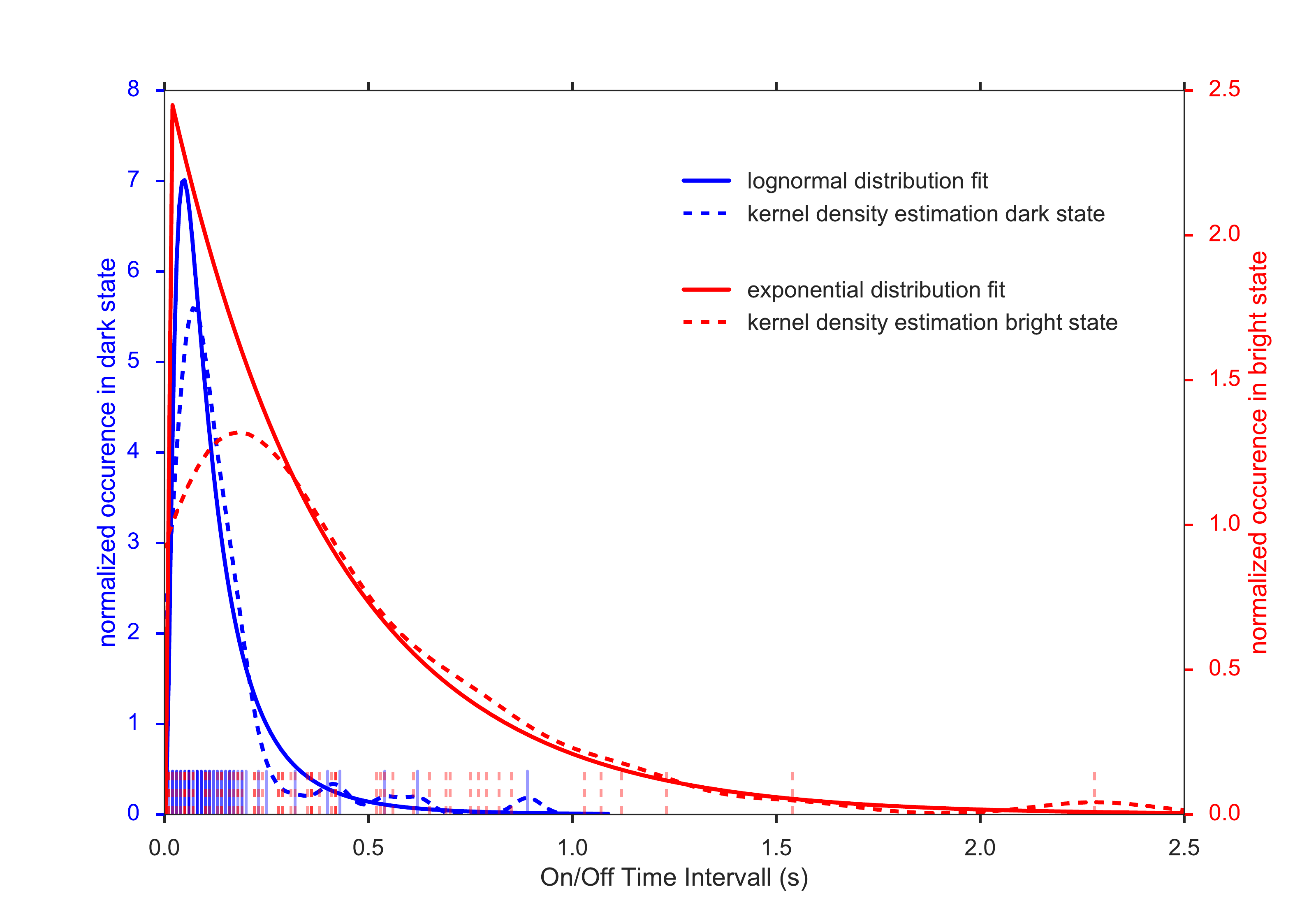}}
					\caption{Time intervals of the single emitter exhibiting the highest blinking rate in the bright (red) and dark (blue) states. Each rising/falling edge of the count rate was individually read out from the time trace. On the horizontal axis small vertical lines represent the individual data points of the bright/dark time intervals. The blue and red dashed curves represent kernel density estimations of the distribution of time intervals of the dark and bright states, respectively. The y-axis is scaled to the normalized kernel density estimate. The red solid line is an exponential fit of the bright state time intervals whereas the blue solid line is a log-normal fit of the dark state time intervals. These fitting functions were chosen because they provide the best agreement with the data using a Kolomogorov-Smirnov test with respect to other functions (p-values: bright state (red) 0.92, dark state (blue) 0.77).}
					\label{fig::fit_blink_distr}
				\end{figure}

				As mentioned in the previous section, the single photon count rates observed from the investigated \sivs varies strongly between a few thousand to a few \SI{100000}{cps}.
				To further investigate the count rate, the luminescence time trajectory of the emitters which exhibit a dip at \gtz is evaluated.
				It is found that some of the observed emitters exhibit fluorescence intermittence, also called blinking (\autoref{fig::blink}).
				Blinking is attributed to temporal ionization of the color center during optical excitation, forming a
charge state which is optically inactive or emitting outside the detection window \cite{Jantzen2016,Neu2012a,Gali2013}.
				Therefore the emitters change between states of higher and lower emission, i.e.\ brighter and darker states, called blinking levels.
				\\
				The photon count time trace of emitter \emnarrow is shown in \autoref{fig::blink}.
				In the overview picture (\autoref{subfig::blink_long}), a few blinking dips can be seen with time intervals of up to a few minutes.
				The fact that the count rate never drops down to the dark count rate lets us assume, that there are at least two \sivs present, one exhibiting fluorescence intermittence and one exhibiting a stable emission.
				When zooming in, shorter time intervals are observable (\autoref{subfig::blink_short}).
				The time intervals range from a few tens of milliseconds up to a few seconds with a few outliers exhibiting very long time intervals up to a few hundred seconds.
				\\
				The bright and dark times exhibit different probability distribution functions and with that, different characteristic time constants.
				In \autoref{fig::fit_blink_distr} the time intervals of the emitter are shown as small vertical dashed red lines and solid blue lines for the bright and dark state respectively.
				Outliers with very long time intervals are ommited here.
				The dashed lines are kernel density estimators of the distribution of the respective time intervals.
				This implies that every data point is represented with a Gaussian function and the resulting functions are added up to model the whole data.
				The red solid line is an exponential fit of the distribution of time intervals in the bright state.
				The high p-value of \num{0.92} confirms the goodness of the fit.
				The median time interval in the bright state obtained by the exponential fit amounts to \SI[separate-uncertainty]{0.09}{s}.
				While other literature on solid state quantum emitters reports an exponential probability distribution for both time intervals in bright and dark states\cite{Bradac2010,Berhane2017}, we found a log-normal probability distribution for the time interval in the dark state.
				The solid blue line in \autoref{fig::fit_blink_distr} is a log-normal fit of the distribution of the time intervals in the dark state.
				A Kolomogorov-Smirnov test yields a p-value of \num{0.77} for the log-normal fit and is by far the best model to describe the data distribution.
				For comparison, the p-value of an exponential fit amounts to \num{0.36}.
				The median time interval in the dark state obtained by the log-normal fit is determined as \SI{0.10}{s}, therefore being close to the median time interval in the bright state.
				Very long time intervals are not shown in the plot for better visualization of the small timescales, however these long time intervals are included in the fit.
				The longest measured time interval amounted to \SI{41.14}{s} and occurred in the dark state.
				Measurements of SiV center blinking time intervals in \cite{Jantzen2016} and \cite{Neu2012a} report time intervals between about \SIrange{0.03}{1}{s}, and \SI{0.1}{s} to \SI{2}{min}, respectively.
				These findings are in good agreement with our measurements. We do not identify a correlation between the count rate of a blinking state and its temporal duration.
				However, a correlation between the position in the bimodal distribution and blinking is established:
				All but one emitters in \hl exhibit blinking, where only one of the emitters in \vl exhibits blinking (\autoref{fig::bimodal_distr}).
				This dependency suggests that emitters in strained nanodiamonds are more likely to exhibit blinking.
				\\
				As blinking is typically linked to temporary loss of photo-excited charges we tentatively explain the observed blinking as a manifestation of the local crystal disorder due to dislocations and impurities which act as a trap for the excited electron and therefore switch the emitter to the dark state \cite{Bradac2010}.
				The assumption that dislocations and impurities are responsible for blinking emitters is in agreement with our findings reported in \ref{subsec::raman}. Regarding the time interval distributions, e.g. research of blinking rhodamine molecules confirmed power law distributed bright state times and log-normal distributed dark state times \cite{Wong2013}. Log-normal distributions are typically explained by a Gaussian distribution of activation barriers of the electron transfer to trap states in the surrounding material \cite{Albery1985} which hints towards a recapture of the electron via multi-phonon relaxation channels.
				\\

		\section{Conclusion}	\label{sec::conclusion}

			In conclusion, in this work we report on a strongly inhomogeneous distribution of \siv spectra in nanodiamonds produced by wet milling from polycrystalline diamond films.
			We find that the zero phonon lines of the emission spectra can be grouped into two clusters:
			\Hl consists of \ZPLs exhibiting a narrow \lw from below \SI{1}{nm} up to \SI{4}{nm} and a broad distribution of \cwl between \SIlist{715;835}{nm}.
			Compared to that, \vl comprises \ZPLs with a broad \lw between just below \SIlist{5; 17}{nm} and \cwl ranging from \SIrange{730}{742}{nm}.
			Based on \emph{ab initio} density functional theory calculations we show that both the observed blue-shifts and red-shifts of the \ZPLs of \vl (as compared to an ideal, unstrained SiV center) are consistently explained by strain in the diamond lattice. 	Further, we suggest, that \hl might be comprised of modified \sivs, the structure of which is currently
unclear.
The broad distribution of emission wavelengths found here covers all earlier results on spectroscopy of SiV centers but considerably extends the range of known emission wavelengths. It further suggests that some single photon emitters in the $715-835$~nm range, previously identified as Cr-, Ni- or Ni/Si-related (see e.g. \cite{Aharonovich2011}), could indeed exist of strained or perturbed SiV centers.
			\\
			Whereas single photon emission could be demonstrated for SiV centers of both clusters, we show that the two clusters of SiV centers show different spectroscopic features: For the phonon sideband spectra we find in
			\vl one prominent peak at a shift of \SI{42}{meV}, which corresponds to a well-known feature assigned to non-localized lattice vibrations \cite{Sternschulte1994,Larkins1971}.
			In \hl we see an accumulation of peaks, at around \SIlist{43;64;150;175}{meV}, which are consistent with sideband peaks reported in \cite{Neu2011,Zaitsev2001,Sternschulte1994}.
						Investigating the time trace of the \siv \pl, we found that predominantly \sivs with narrowband emission (\hl) exhibit fluorescence intermittence with on/off times between several microseconds up to \SI{41}{s}.
			Furthermore, we see an exponential distribution of bright time intervals and a log-normal distribution of dark time intervals, consistent with research on single molecules \cite{Wong2013}.
\\
In summary, whereas SiV centers produced by ion implantation in high quality bulk diamond material show very reproducible spectral properties, SiV centers in nanodiamonds produced by milling techniques or CVD growth feature strongly varying optical spectra. This, on one hand, limits their applicability for quantum information tasks requiring indistinguishable emitters and, on the other hand, demonstrates the need for development of low-strain, nanometer-sized diamond material with low defect density.

\ack

We thank Elke Neu, Saarland University, for helpful discussions, Martin Kamp, University of Wuerzburg, for support with some of the TEM images and Matthias Schreck, University of Augsburg, for providing Ir-coated substrates. This work was partially funded by the European Union 7th Framework Program under Grant Agreement No. 61807 (WASPS) and Deutsche Forschungsgemeinschaft (FOR1493). AG acknowledges the National Research Development and Innovation Office of Hungary within the Quantum Technology National Excellence Program (Project Contract No. 2017-1.2.1-NKP-2017-00001).


	\bibliographystyle{iopart-num}
    	\bibliography{library}

\end{document}